\documentclass[manuscript=article, journal=mamobx, layout=twocolumn]{achemso}
\setkeys{acs}{doi = false}
\renewcommand\tocentryname{}

\makeatletter
\renewcommand*{\acs@tocentry@print}[1]{%
    \gdef\acs@tocentry@text{\normalsize#1}%
    \acs@tocentry@print@aux%
}
\renewcommand*{\acs@tocentry@print@aux}{%
    \begingroup 
    \let\@startsection\acs@startsection@orig 
    \singlespacing 
    \acs@section*{\tocentryname}%
    \vspace{-2.0em}
    \centering
    \begin{center} 
            {%
                \begin{minipage}{\acs@tocentry@width} 
                \vbox to \acs@tocentry@height{\acs@tocentry@text}%
                \end{minipage}%
            }%
    \end{center}%
    \endgroup 
}
\makeatother

\let\oldmaketitle\maketitle
\let\maketitle\relax

\usepackage{chemformula}
\usepackage{subcaption}
\usepackage{seqsplit}

\newcommand{\angstrom}{\textup{\AA}}
\newcommand*{\br}[1][]{\mathbf{r}_{#1}}                  
\newcommand*{\brsup}[1][]{\mathbf{r}^{#1}}               
\newcommand*{\bR}{\mathbf{R}}                            
\newcommand*{\dr}[1][]{\mathrm{d}\mathbf{r}_{#1}\ }      
\newcommand*{\drsup}[1][]{\mathrm{d}\mathbf{r}^{#1}\ }   
\newcommand*{\dvar}[2][]{\mathrm{d #2}_{#1}\ }           
\author{Takashi J. Yokokura}
\affiliation[University of California Berkeley]
{Department of Chemical and Biomolecular Engineering, University of California Berkeley, Berkeley, California 94720, USA}
\author{Chao Duan}
\affiliation[University of California Berkeley]
{Department of Chemical and Biomolecular Engineering, University of California Berkeley, Berkeley, California 94720, USA}
\author{Erika A. Ding}
\affiliation[University of California Berkeley]
{Department of Chemical and Biomolecular Engineering, University of California Berkeley, Berkeley, California 94720, USA}
\author{Sanjay Kumar}
\affiliation[University of California Berkeley]
{Department of Chemical and Biomolecular Engineering, University of California Berkeley, Berkeley, California 94720, USA}
\alsoaffiliation[University of California Berkeley]
{Department of Bioengineering, University of California, Berkeley, CA 94720, USA}
\author{Rui Wang}
\email{ruiwang325@berkeley.edu}
\affiliation[University of California Berkeley]
{Department of Chemical and Biomolecular Engineering, University of California Berkeley, Berkeley, California 94720, USA}
\alsoaffiliation[Lawrence Berkeley National Lab]
{Materials Sciences Division, Lawrence Berkeley National Lab, Berkeley, California 94720, USA}

\title{Effects of Ionic Strength on the Morphology, Scattering, and Mechanical Response of Neurofilament-derived Protein Brushes}


\begin{document}


\twocolumn[
\begin{@twocolumnfalse}
\oldmaketitle
\begin{abstract}
Protein brushes not only play a key role in the functionality of neurofilaments but also have wide applications in biomedical materials. 
Here, we investigate the effect of ionic strength on the morphology of protein brushes using a continuous-space self-consistent field theory. 
A coarse-grained multi-block charged macromolecular model is developed to capture the chemical identity of amino acid sequences. 
For neurofilament heavy (NFH) brushes at pH 2.4, we predict three morphological regimes: swollen brushes, condensed brushes, and coexisting brushes which consist of both a dense inner layer and a diffuse outer layer. 
The brush height predicted by our theory is in good agreement with experimental data for a wide range of ionic strengths. 
The dramatic height decrease is a result of the electrostatic screening-induced transition from the overlapping state to the isolated state of the coexisting brushes. 
We also study the evolution of the scattering and mechanical responses accompanying the morphological change. The oscillation in the reflectivity spectra characterizes the existence and microstructure of the inner condensed layer, whereas the shoulder in the force spectra signifies the swollen morphology. 
\nolinebreak
\begin{tocentry}
    \includegraphics{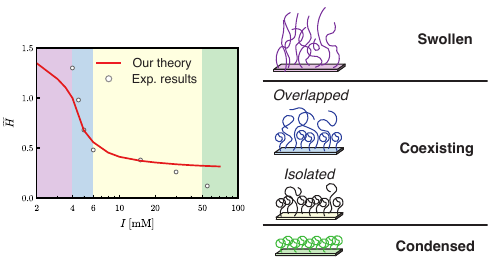}
\end{tocentry}

\end{abstract}
\end{@twocolumnfalse}
]

\clearpage
\section{Introduction}
Neurofilaments (NFs) are cylindrical, self-assembled protein filaments organized axially within the axon. 
Brushes consisting of intrinsically disordered proteins (IDPs) protrude out from the NF cores, which play a critical role in the stability, organization, and functionality of the neuron \cite{yuan_neurofilaments_2017, zhu_delayed_1997}. 
Mutations in the comprising proteins have also been linked to neurodegenerative diseases such as amyotrophic lateral sclerosis, Alzheimer's disease, and Parkinson's disease, in which the instability of the NFs may contribute to neuronal cell death \cite{yuan_neurofilaments_2012, barro_blood_2020, lavedan_mutation_2002}.
Furthermore, protein-inspired brushes arouse great interest in practical applications because of their unique properties in biocompatibility \cite{langer_designing_2004, pan_interfacial_2020}, sensitivity to external stimuli \cite{das_polyelectrolyte_2015, conrad_towards_2019}, and ease of genetic modification \cite{blum_stimuli_2022}. These advantages enable their wide application in biomedical devices and materials, such as sensors \cite{wiarachai_clickable_2016, xie_chiral_2015, welch_polymer_2011}, valves \cite{ito_signal-responsive_2000, yameen_single_2009}, actuators \cite{kelby_controlled_2011, ionov_reversible_2006}, artificial cartilage \cite{kobayashi_tribological_2010}, and vehicles for drug/gene delivery \cite{liechty_polymers_2010}.

Understanding the morphology and interactions of protein and protein-inspired brushes remains a great challenge. In an earlier work, the Kumar group modified the tail domain of neurofilament heavy (NFH), which is a major component of neurofilaments, and further grafted the recombinant protein to a planar surface \cite{srinivasan_stimuli-sensitive_2014}. They observed a dramatic height change in response to the addition of monovalent salt at a critical ionic strength: the height reduces by a factor of three within a narrow salt concentration range of 2 mM. This height reduction is far steeper than the $-1/3$ scaling predicted by existing theories for homogeneous polyelectrolyte brushes \cite{israels_charged_1994, chen_50th_2017}. The height was also found to be sensitive to pH. 
Additionally, there is debate on the role of NFH in structurally stabilizing the axon. Whether this stability is attributed to steric effects as a result of cross-links between neighboring NFHs or induced by electrostatic repulsion is not clear \cite{fuchs_structural_1998, zhulina_polymer_2010}. 
There is additional controversy on whether NFH plays a role in setting the axonal diameter based on transgenic mouse studies \cite{elder_requirement_1998, zhu_disruption_1998, rao_neurofilament-dependent_1998}.
Besides external stimuli, protein brush morphology also depends on the intrinsic chemical properties encoded by the amino acid sequence of the comprising proteins. 
By replacing serine with more charged aspartate residues, the Kumar group found that the backbone charge density of the protein has a significant effect on the height of NFH-inspired brushes\cite{bhagawati_site-specific_2016}. 
They also found that brush height is heavily influenced by protein phosphorylation \cite{lei_structural_2018}.
Similar morphological behaviors have also been observed in synthetic polyelectrolyte brushes \cite{hest_protein-based_2001, chen_50th_2017}.

Great efforts have been made to theoretically study the morphological behaviors of protein-inspired brushes using polymer brush models. While the pioneering scaling theory can predict macroscopic information such as brush height, it is unable to describe the density distribution of the constituent monomers \cite{alexander_adsorption_1977, de_gennes_conformations_1980}. Furthermore, the so-called “classical” or “strong stretching theory” is restricted to highly stretched chains due to the classical path approximation \cite{milner_theory_1988, milner_s_t_polymer_1991, skvortsov_structure_1988}. It fails to capture brushes with other chain conformations and cannot explicitly account for the amino acid sequence. Recently, lattice self-consistent field theory (commonly known as the Scheutsjen-Fleer model) \cite{wijmans_self-consistent_1992, scheutjens_statistical_1979, scheutjens_statistical_1980, leermakers_modeling_2010}, has been widely applied to describe synthetic and protein-inspired brushes. Using this method, \citeauthor{zhulina_effect_2007} predict the morphological response of NFs to ionic strength and pH \cite{zhulina_effect_2007, zhulina_self-consistent_2007}. However, the experimentally observed dramatic height change of NFH brushes with the addition of salt at low pH has not been investigated by the lattice self-consistent field theory \cite{srinivasan_stimuli-sensitive_2014}.

\begin{figure*}[hbt!]
\centering{
\phantomsubcaption\label{fig:brush}
\phantomsubcaption\label{fig:coarse}
\phantomsubcaption\label{fig:disc}
}
\includegraphics[width=1.00\textwidth]{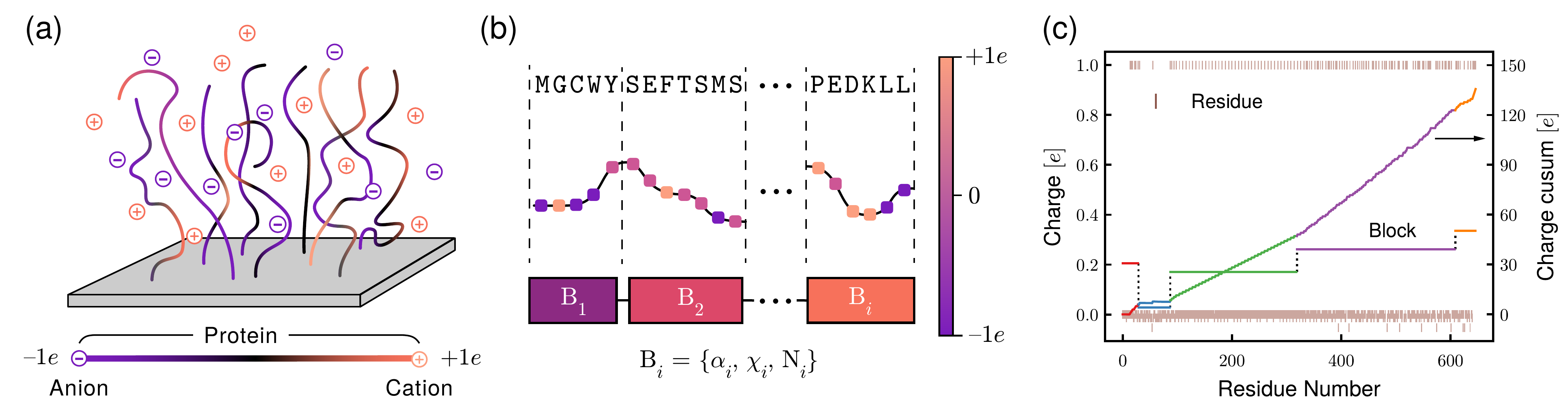}
\caption{
(a) Schematic of IDP brush immersed in salt solution. Colorbar illustrates the charge of different species. 
(b) Schematic of the coarse-graining approach to represent an IDP with a particular amino acid sequence by the multi-block charged macromolecular model. Each block $B_i$ is parameterized by its backbone charge density $\alpha_i$, hydrophobicity $\chi_i$, and equivalent number of Kuhn segments $N_i$.
(c) Application of the coarse-graining procedure to NFH at pH 2.4. The charge of each amino acid residue, cumulative sum (cusum), and block charge distribution are plotted against the residue number. 
}
\label{fig:scheme}
\end{figure*}

Characterizing the microstructure of protein brushes is critical to understanding their morphological response, surface activity, and functionality. However, the overall height as obtained from spectroscopic ellipsometry or atomic force microscopy (AFM) measurements cannot directly provide the detailed distribution of amino acids in the brush. In this work, we apply a continuous-space self-consistent field theory (SCFT) to elucidate the relationship between the chemical structure and morphological behavior of protein brushes. A coarse-graining approach built upon a multi-block charged macromolecular model is developed to capture the chemical identity of the amino acid sequence. We investigate the morphological response of NFH brushes to ionic strength. The theoretical prediction shows good agreement with the experimental results reported in the literature \cite{srinivasan_stimuli-sensitive_2014}. The evolution of scattering spectra and mechanical properties accompanying the morphological changes is also examined. We show the effectiveness of reflectivity and force spectroscopy in detecting the microstructure of protein brushes.

\section{Model and Theory}
We consider a system consisting of intrinsically disordered proteins (IDPs) grafted on a planar surface as shown in Fig.~\ref{fig:brush}. The system is treated as a semi-grand canonical ensemble: the number of proteins is fixed, whereas solvent and mobile ions are connected with a bulk salt solution of ion concentration $c_\pm^{b}$ to maintain the chemical potentials of solvent $\mu_s$ and ions $\mu_\pm$. Mobile ions are taken as point charges with valency $z_\pm$. The widely adopted multi-block charged macromolecular model \cite{zhulina_self-consistent_2007, bianchi_relevance_2020, lee_effects_2013} is used to represent IDPs. The charged macromolecules are assumed to be Gaussian chains of $N$ total segments with Kuhn length $b$. This is also a general model for synthetic polyelectrolytes and other biomacromolecules \cite{choi_physical_2020, bianchi_relevance_2020}.

We develop a coarse-graining approach to map the chemical identity of any amino acid sequence to the multi-block charged macromolecule. Adjacent amino acids with similar charge and hydrophobicity are grouped into a block as illustrated in Fig.~\ref{fig:coarse}. Each block is thus a section of homopolymer with number of residues $N_i^*$, smeared backbone charge density $\alpha_i$, and hydrophobicity $\chi_i$ (manifested in terms of the Flory-Huggins parameter accounting for the short-range van der Waals interactions between amino acids and solvent). The equivalent number of Kuhn segments in this block is $N_i = N_i^* l_A / b$, where $l_A=0.36$ nm is the average contour length of an amino acid \cite{dietz_protein_2006, oesterhelt_unfolding_2000}. In this work, the grouping procedure is simplified by only considering the charge distribution along the amino acid sequence. Each amino acid is treated as a weak acid/base, where its charge is calculated from the dissociation constant ($\mathrm{pK_a}$/$\mathrm{pK_b}$) of its residue and the bulk pH through the Henderson-Hasselbalch equation \cite{lide_crc_1991}. An alternative but more rigorous treatment can be achieved by explicitly considering the effect of local proton concentration on the dissociation equilibrium (See Sec. II.2 of the SI for the annealed model for protein charge). The blocks are determined by tracking the cumulative sum (cusum) of the charge distribution along the amino acid sequence. The section of $N_i^*$ amino acids which contributes to a nearly constant slope in the cusum is grouped as a block. The charge density of this block $\alpha_i$ is directly identified by the slope. 

Fig.~\ref{fig:disc} illustrates the coarse-graining procedure for NFH at pH 2.4, from the initial charge of each amino acid residue to the cusum and finally to the length and charge of each block. Because amino acids are chemically similar, we assume that the hydrophobic interactions between different blocks are negligible. To determine the block hydrophobicity $\chi_i$, residues are categorized into apolar, polar, and charged groups as used in previous work \cite{zhulina_effect_2007}. Each group is assigned a corresponding Flory-Huggins parameter based on the values reported in the literature \cite{zhulina_effect_2007}. The hydrophobicity $\chi_i$ of each block is then the average of its constituent amino acid Flory-Huggins parameters. The resulting sets of $\alpha_i$ and $\chi_i$ for NFH are presented in Table~\ref{tab:NFH}. The details of the NFH sequence are provided in Sec. I of the SI. The numerical results obtained by the theory are found not to be sensitive to the fineness of the coarse-graining model as long as the number of blocks is large enough (see the sensitivity analysis in Sec. III of the SI).

\begin{table}[hbt!]
\caption{Partitions of amino acid residues, charge density $\alpha_i$, and hydrophobicity $\chi_i$ of coarse-grained blocks for NFH at pH 2.4 in the multi-block charged macromolecular model.}
\label{tab:NFH}
\centering
    \begin{tabular}{c l c c}
    \hline\hline
    \noalign{\smallskip}
    Block & Residues  & $\alpha_i$ & $\chi_i$\\
    \hline\noalign{\smallskip}
    \textbf{1} & [1,~~~28] & 0.204967 & 1.586207\\
    \textbf{2} & [29,~~87] & 0.027801 & 1.434483\\
    \textbf{3} & [88,~~319] & 0.170493 & 2.113793\\
    \textbf{4} & [320,~609]& 0.261110 & 1.534483\\
    \textbf{5} & [610,~647]& 0.336030 & 0.989474\\
    \noalign{\smallskip}
    \hline\hline
    \end{tabular}
\end{table}

The semi-grand canonical partition function can be written as:
\begin{align}
\Xi =~&\frac{1}{n_p!\nu^{N n_p}} \prod_{\gamma}\sum_{n_\gamma=0}^{\infty} \frac{e^{\mu_\gamma n_\gamma}}{n_\gamma!v_\gamma^{n_\gamma}}\label{eqn:Xi}\\ 
&\prod_{i=1}^{n_p} \int \mathcal{D}\{\bR_i\}\int \prod_{j = 1}^{n_\gamma} \dr[\gamma, j] \exp \bigg\{-\mathcal{H} \bigg\}\nonumber\\ 
&\prod_{\br} \delta(\hat{\phi}_p(\br) + \hat{\phi}_s(\br) -1)~,\nonumber
\end{align}
where $\gamma=s, \pm$ represents the solvent, cations, and anions, respectively. $\nu_p$ is the volume of each segment in the $n_p$ proteins, while $\nu_\gamma$ is the volume of each of the $n_\gamma$ small molecules. For simplicity, we assume $\nu_p = \nu_s = \nu$. $\mathcal{D}\{\bR_i \}$ denotes the integration over all chain configurations of protein $i$. $\hat{\phi}_p(\br)$ and $\hat{\phi}_s(\br)$ are the local instantaneous volume fraction of the protein and solvent, respectively. The $\delta$ function at the end of Eq.~\ref{eqn:Xi} accounts for the incompressibility. The Hamiltonian $\mathcal{H}$ is given by: 

\begin{align}
\mathcal{H} =~&\sum_{k=1}^{n_p} \frac{3}{2b^2} \int_{0}^{N} \dvar{s} \bigg(\frac{\partial \bR_k(s)}{\partial s}\bigg)^2\label{eqn:H}\\ &+ 
\frac{1}{\nu} \int \dr \sum_{i=1}^{\xi}\bigg(\chi_i \hat{\phi}_i(\br)\hat{\phi}_s(\br)\nonumber\\ 
&\quad\quad\quad\quad\quad\quad\quad+ \sum_{j\ne i}^{\xi} \frac{\chi_{ij}}{2}\hat{\phi}_i(\br)\hat{\phi}_j(\br)\bigg)\nonumber\\
&+ \frac{1}{2} \int \dr \drsup[\prime]\ \hat{\rho}_c(\br)\ C(\br, \br^\prime)\ \hat{\rho}_c(\br^{\prime})\nonumber
\end{align}
which consists of three contributions: (1) the intra-chain elastic energy of IDPs, (2) the short-range hydrophobic interactions between blocks and solvents as well as cross-interactions between blocks, and (3) the long-range Coulomb interactions between charged species. $\hat{\phi}_i(\br)$ is the instantaneous volume fraction of block $i$, where $\xi$ is the total number of blocks in each IDP. Thus, $\hat{\phi}_p(\br) = \sum_{i=1}^{\xi} \hat{\phi}_i(\br)$. $\chi_i$ is the Flory-Huggins parameter between block $i$ and solvents, whereas $\chi_{ij}$ is the parameter between blocks $i$ and $j$. Here, we neglect the hydrophobic interactions between different blocks, i.e., $\chi_{ij} = 0$. Furthermore, $\hat{\rho}_c(\br)  = z_+\hat{c}_+(\br) - z_-\hat{c}_-(\br) + \sum_{i=1}^{\xi} \alpha_i \hat{\phi}_i(\br)/\nu$ is the local charge density, where $\hat{c}_\pm(\br)$ is the instantaneous number of ions. $C(\br, \br^{\prime})$ is the Coulomb operator satisfying $-\nabla\cdot\big[\epsilon(\br)\nabla C(\br, \br^{\prime})\big]=\delta(\br- \br{^\prime})$. $\epsilon(\br) = kT \epsilon_0\epsilon_r(\br)/e^{2}$ is the scaled permittivity, where  $\epsilon_0$ is the vacuum permittivity, $e$ is the elementary charge, and $\epsilon_r(\br)$ is the local dielectric constant which depends on the local composition of the system. 

We follow the standard self-consistent field procedure \cite{fredrickson_equilibrium_2006, duan_conformation_2020, duan_association_2022, duan_stable_2022, duan_electrostatics-induced_2023} (see Sec. II in the SI for the detailed derivation). First, the interacting system is decoupled into non-interacting proteins and ions in fluctuating fields by identity transformations and the Hubbard-Stratonovich transformation. Next, the functional integral over the fluctuating fields is replaced by the saddle-point approximation. The resulting self-consistent equations for block density $\phi_i$, electrostatic potential $\psi$, and conjugate fields $w_i$ and $w_s$ are: 
\begin{subequations} \label{SCF:main}
\begin{align}
&w_i(\br) - w_s(\br) = \chi_i(1-\phi_p(\br))-\alpha_i\psi(\br)\label{SCF:a}\\
&\quad\quad\quad\quad-\frac{\partial\epsilon(\br)}{\partial\phi_i}\frac{\lvert \nabla \psi(\br) \rvert^2}{2}\nu -{\textstyle\sum_{i=1}^{\xi}} \chi_{i}\phi_i(\br)\nonumber\\ 
&\phi_{i}(\br) = \frac{n_p}{Q_p} \int_{\sum_{j=0}^{i-1}N_{j}}^{\sum_{j=0}^{i}N_j} \mathrm{ds}\ q(\br;\mathrm{s}) q_c(\br;\mathrm{s})\label{SCF:b}\\
&1-\phi_p(\br) = e^{\mu_s}\exp(-w_s(\br))\label{SCF:c}\\
&-\nabla\cdot \big(\epsilon(\br)\nabla\psi(\br) \big)=\label{SCF:d}\\ 
&\quad\quad\quad\quad z_+c_+(\br)-z_-c_-(\br)  +{\textstyle\sum_{i=1}^{\xi}}\frac{\alpha_i}{\nu} \phi_i(\br)\nonumber
\end{align}
\end{subequations}
$c_\pm = \lambda_\pm\exp(\mp z_\pm\psi)$ is the ion concentration, where $\lambda_\pm = e^{\mu_\pm}/\nu_\pm $ is the fugacity of the ions determined by the bulk ion concentration $c_\pm^b$. $Q_s = \nu^{-1}\int \dr \exp{(-w_s)}$ is the single-particle partition function of the solvent. $Q_p = \nu^{-1} \int \dr q(\br;\mathrm{s})$ is the single-chain partition function of the protein. $q(\br; \mathrm{s})$ is the propagator which satisfies the modified diffusion equation:
\begin{align}
\frac{\partial}{\partial s} q(\br;&\mathrm{s}) = \frac{b^2}{6}\nabla^2 q(\br;s)- w_i(\br)q(\br;s)\;,\label{eqn:MDE}\\
\text{where}\; &w_i(\br) =\nonumber\\ 
&\begin{cases}
w_{1}(\br) & \text{for } s = [0, N_1]\\
&\vdots \\
w_{\xi}(\br) & \text{for } s = [\sum_{j=0}^{\xi-1}N_j, \sum_{j=0}^{\xi} N_j]
\end{cases}\nonumber
\end{align}
While our theory is general for any geometry of the system, here we consider a one-dimensional planar system, where the protein densities vary only in the $z$-direction but remain homogeneous in the $xy$-plane. The initial condition for the tethered end of the protein is thus $q(\mathrm{z};0)=\delta(z-z^*)$, where $z^*\to 0+$. $q_c(\mathrm{z};s)$ in Eq.~\ref{SCF:b} is introduced as the complementary propagator starting at the free end which also satisfies the same modified diffusion equation (Eq.~\ref{eqn:MDE}) as $q(\mathrm{z};s)$ but with the initial condition $q_c(\mathrm{z},N)=1$. We assume that there is no interaction between proteins and the substrate. The boundary condition corresponding to this non-interacting surface is to set the protein density of each block $\phi_i$ to zero. The gradient of the electrostatic potential is also set to zero on the substrate. The free energy per unit area is:
\begin{align}
F =&-\sigma \ln Q_p - e^{\mu_s}Q_s\label{eqn:F}\\ 
&+\frac{1}{\nu}\int \dvar{z} \bigg[ \sum_{i=1}^{\xi} \bigg(\chi_i\phi_i (1-\phi_p) - w_i\phi_i\bigg)\nonumber\\ 
&- w_s(1-\phi_p) \bigg] + \int \dvar{z} \bigg[-\frac{\epsilon}{2}\lvert \nabla\psi \rvert^2\nonumber\\ 
&\quad\quad\quad\quad\quad\quad\quad+\frac{\psi}{\nu}\sum_{i=1}^{\xi}\alpha_i\phi_i - c_+-c_-\bigg]~,\nonumber
\end{align}
where $\sigma$ denotes the grafting density of proteins on the surface. 

The equilibrium protein density profile, electrostatic field, and ion distribution can be obtained by solving Eq.~\ref{SCF:main} and Eq.~\ref{eqn:MDE} iteratively. The height of the brush $H$ is extracted from the protein density profile based on the Gibbs dividing surface \cite{israelachvili_intermolecular_2011, wang_theory_2014}:
\begin{equation}\label{eq:H}
H = \frac{2\int_{0}^{\infty} z\phi_p(\mathrm{z})\dvar{z} }{\int_{0}^{\infty}  \phi_p(\mathrm{z})\dvar{z}}
\end{equation}
Compared to the lattice constraint invoked in previous work \cite{scheutjens_statistical_1979, scheutjens_statistical_1980, zhulina_self-consistent_2007}, the differential equations are solved in the continuous space, where the numerical discretization is decoupled from the physical lattice. This improves the flexibility of the calculations \cite{chantawansri_spectral_2011}.
An iterative, centered finite difference scheme is used to solve the Poisson-Boltzmann equation (Eq.~\ref{SCF:d}), whereas the Crank-Nicolson scheme \cite{hoffman_numerical_2018} is used to solve the modified diffusion equation (Eq.~\ref{eqn:MDE}). Our theory can be easily generalized to consider brushes with protein mixtures, various chain architectures, and morphologies with inhomogeneity in multiple dimensions.
\section{Results and Discussion}
The theory presented above can be applied to any amino acid sequence. Here, we focus on the effect of ionic strength on a planar brush comprised of modified NFH sidearms at pH 2.4. According to this bulk proton concentration, the ionic strength ($I=\frac{1}{2}\sum_i z_i^2 c_i$ for ion species $i$) for the salt-free system is thus 4 mM. 
For consistency with the setup in the experiments of the Kumar group \cite{srinivasan_stimuli-sensitive_2014}, we consider the same grafting density $\sigma = 0.02\ \mathrm{nm}^{-2}$ and the addition of monovalent salt ($z_\pm = 1$). For simplicity, the dielectric constant of the system is assumed to be uniform and set to be $\epsilon_{r}= 80$, the value of water. The temperature is set at 293 K, yielding a Bjerrum length $l_B = e^2/4\pi kT\epsilon_0 \epsilon_r$ of $0.7$ nm. 

\subsection{Morphological Response to Ionic Strength}

\begin{figure}[b!]
    \centering
    \includegraphics[width=0.95\linewidth]{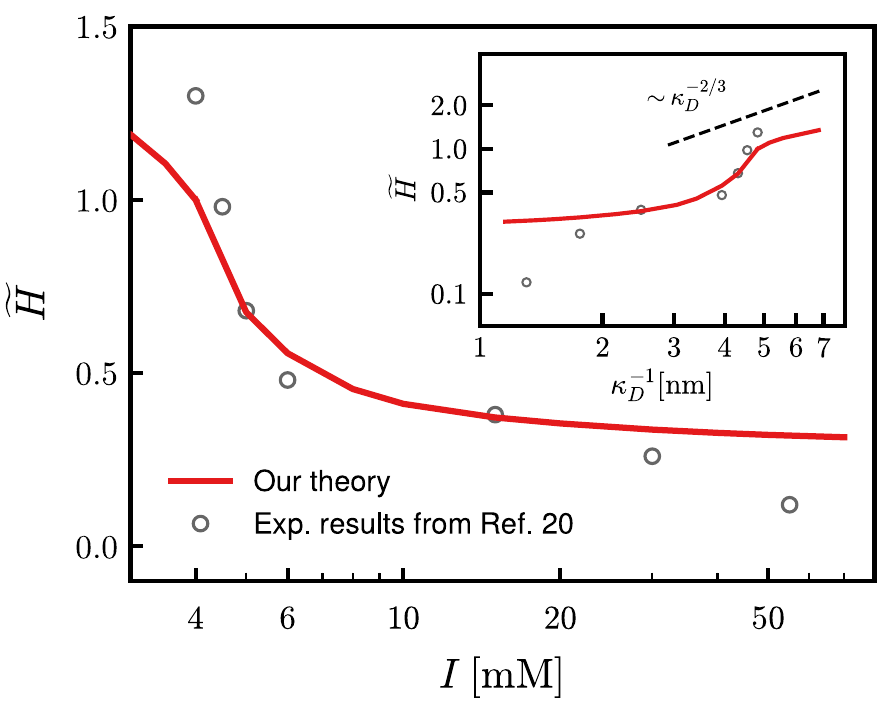}
    \caption{Height response of NFH brushes at pH 2.4 to various ionic strengths $I$. $\tilde{H}$ is the normalized brush height defined using the value under the salt-free condition as a reference. The theoretical prediction (solid line) is compared to the experimental results (circles) reported by the Kumar group \cite{srinivasan_stimuli-sensitive_2014}. The inset plots $\tilde{H}$ as a function of the Debye screening length $\kappa_D^{-1}$ to illustrate the scaling behavior at high $\kappa_D^{-1}$. $b=3.00\ \mathrm{nm}$ and $\nu =1.30\ \mathrm{nm^3}$ are adopted for the best fit to experimental data.}
    \label{fig:prof}
\end{figure}

The morphology of NFH brushes is determined by the interplay between the hydrophobic interaction, screened electrostatic repulsion, and conformational entropy of grafted proteins. 
Fig.~\ref{fig:prof} shows that, after properly choosing model parameters $b$ and $\nu$, the height response predicted by our theory is in good agreement with the experimental results reported in the literature for NFH brushes at 2.4 pH over a wide range of ionic strengths \cite{srinivasan_stimuli-sensitive_2014}.
Here, the brush heights are normalized by the value under the salt-free condition (i.e., $H = 33$ nm at $I=4$ mM) for a better comparison to the experimental values. We note that the experimental heights measured by the Kumar group were normalized by 57 nm \cite{srinivasan_stimuli-sensitive_2014}. The difference of the brush height in the salt-free case between experiment and theory could be attributed to the methods for quantifying the brush height. The experimental heights were determined using atomic force microscopy, while our theoretical work uses the Gibbs dividing surface, as defined in Eq.~\ref{eq:H}.
The height response of NFH brushes to increasing ionic strength can be divided into three regimes. At very low ionic strengths ($I < 4$ mM or $\kappa_D^{-1} > 4.8$ nm, where $\kappa_D^{-1}$ is the Debye screening length defined as $\kappa_D^{-1} = (8\pi l_B I)^{-1/2}$), the normalized height $\tilde{H}$ decreases as $I$ increases. The theoretical results follow the scaling of $\tilde{H} \sim \kappa_D^{-2/3}$ as shown in the inset of Fig.~\ref{fig:prof}, consistent with the picture of the Alexander-de Gennes model for strongly stretched brushes. In the intermediate regime of $4$ mM $< I < 6$ mM, the brushes collapse dramatically when a small amount of monovalent salt is added. $\tilde{H}$ decreases by a factor of 3 within a narrow ionic strength range of 2 mM, which fully captures the dramatic height change observed in experiments. From $I = 6$ mM onwards, the height decrease slows until the charges carried by the proteins are completely screened and the brush morphology eventually approaches that of a condensed, charge-neutral brush.

We note that there are some discrepancies between theory and experiments at very low and very high ionic strengths. This may be attributed to the simplified treatment of the dielectric environment in the current work, where the dielectric permittivities of the solvent, protein, and substrate are not distinguished. At low ionic strength, the image force due to the dielectric contrast between the substrate and solvent will repel ions further away from the surface, leading to less screening and higher brush height. At the other limit of high ionic strength, proteins form a dense layer near the surface. Since the dielectric constants of proteins are usually much lower than that of water, the strength of the ionic screening ($\kappa \sim (2I/\epsilon_r)^{1/2}$) is higher than the theoretical prediction, giving rise to a lower brush height.

\begin{figure}[hbt!]
    \centering
    \includegraphics[width=0.85\linewidth]{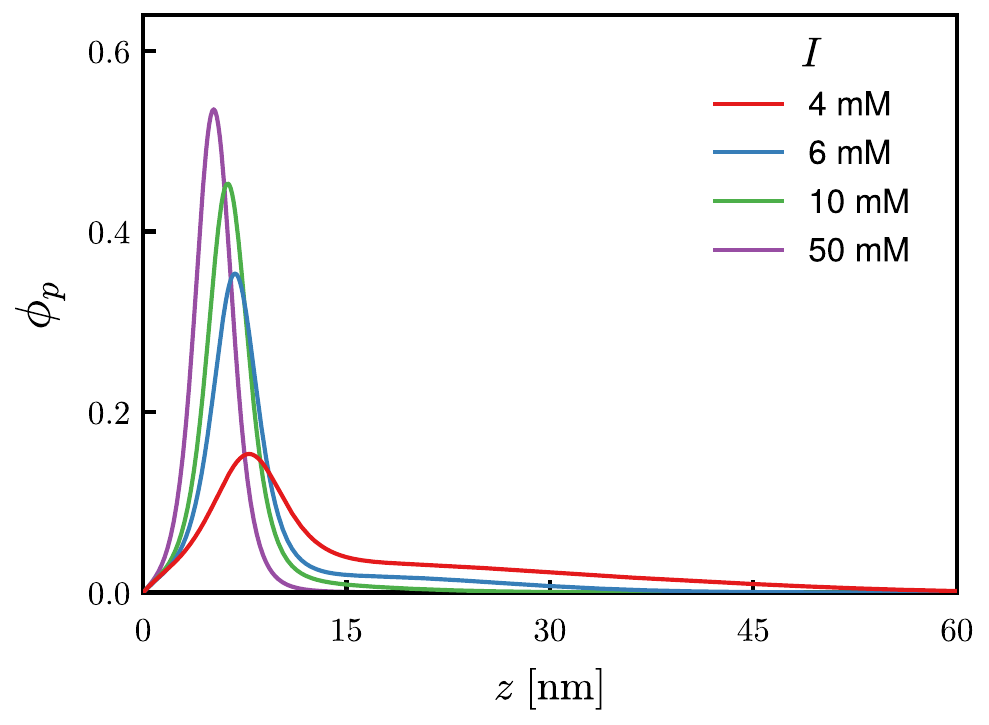}
    \caption{Representative distributions of the overall density of NFH proteins $\phi_p$ for $I = 4, 6, 10,$ and $50$ mM. $z$ denotes the direction perpendicular to the grafting surface.}
    \label{fig:phas}
\end{figure}

Fig.~\ref{fig:phas} depicts the representative protein density profiles underlying the overall brush height. The corresponding ionic density profiles are provided in Sec. IV of the SI.
At low ionic strengths (e.g., $I = 4$ mM), the intrachain electrostatic repulsion is dominant. Each protein adopts a strongly stretched conformation which leads to a swollen brush morphology. The protein density distribution is quite diffuse with a pronounced tail stretching into the solution. The peak is attributed to the aggregation of less charged residues near the grafting point. 
In stark contrast, at high ionic strength (e.g., $I = 50$ mM), the charges on the protein are largely screened, whereas hydrophobic attraction between residues is dominant. Each protein chain collapses and thus the whole brush adopts a condensed morphology. This is reflected by the single sharp peak in the density profile. 
Furthermore, at intermediate ionic strengths between the two limiting regimes, e.g., $I = 6$ and $10$ mM, brushes exhibit characteristics of both the swollen morphology and the condensed morphology: there is a coexistence between a diffuse outer layer and a dense inner layer. In fact, the dramatic height change from 4 mM to 6 mM originates from the morphological transition from a swollen brush to a coexisting brush.

\begin{figure}[hbt!]
    \centering{\phantomsubcaption\label{fig:block3} \phantomsubcaption\label{fig:block5}\phantomsubcaption\label{fig:blocki_schem}}
    \includegraphics[width=0.85\linewidth]{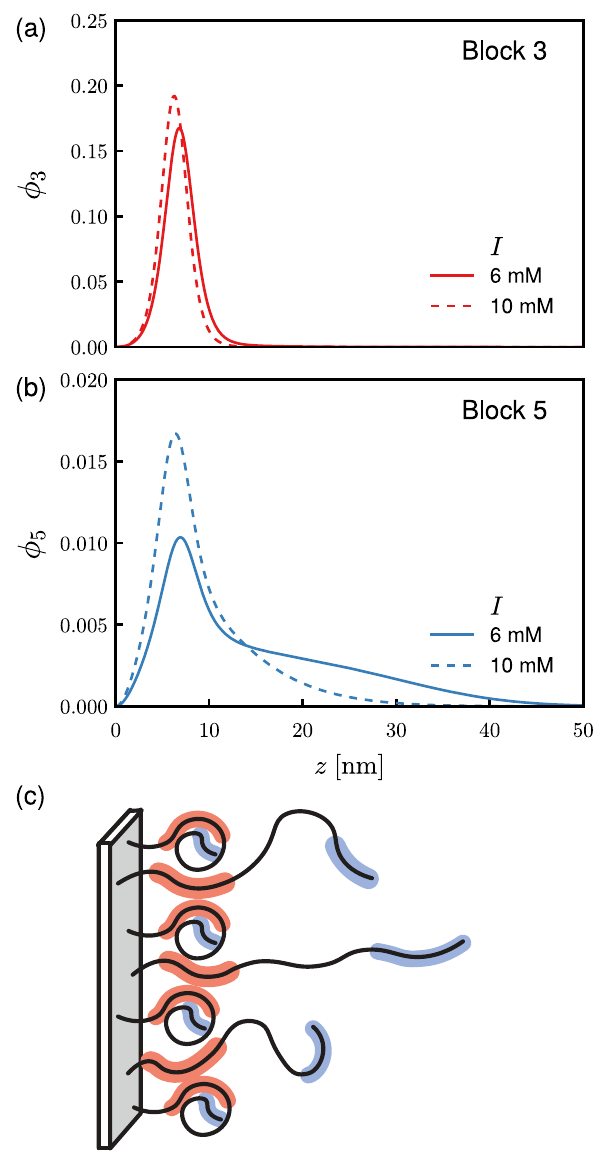}
    \caption{The morphology of coexisting brushes. The density distribution of (a) Block 3 and (b) Block 5 at $I = 6$ and 10 mM. (c) Schematic of the coexisting brush morphology with two populations of chain conformations. Block 3 and Block 5 are highlighted by red and blue, respectively, for illustration.}
    \label{fig:blocki}
\end{figure}

Our theory facilitates the calculation of the density distribution of each constituent protein residue, which enables us to further elucidate the microstructure of the coexisting morphology. In our multi-block charged macromolecular model, this corresponds to the density distribution of a particular block as indicated by Eq.~\ref{SCF:b}. 
Fig.~\ref{fig:blocki} shows the distribution of two representative blocks at $I = 6$ and $10$ mM. Block 3 (Fig.~\ref{fig:block3}) is the middle block less affected by the grafted chain end and is moderately charged. Block 5 (Fig.~\ref{fig:block5}) is the ending block which contains the free chain end and is the most positively charged among all the blocks.
As shown in Fig.~\ref{fig:blocki}, these two blocks exhibit remarkably different responses to changing $I$. Block 3 remains collapsed in the inner condensed layer. Its density distribution is largely insensitive to $I$, only with a slight compression at 10 mM due to the increase of electrostatic screening. 
On the other hand, the distribution of Block 5 is bimodal, signifying its presence both in the inner condensed layer and in the outer diffuse layer. In stark contrast to Block 3, the density distribution of Block 5 changes significantly with $I$. As $I$ increases from 6 mM to 10 mM, a pronounced fraction of residues move from the outer diffuse layer to the inner condensed layer. The density of remaining residues in the outer layer shrinks significantly, which is reflected by a decrease of the brush height as shown in Fig.~\ref{fig:prof}.       

\begin{figure}[hbt!]
\centering{
\phantomsubcaption\label{fig:trans_a}
\phantomsubcaption\label{fig:trans_b}
\phantomsubcaption\label{fig:trans_c}
\phantomsubcaption\label{fig:trans_d}
}
\includegraphics[width=0.85\linewidth]{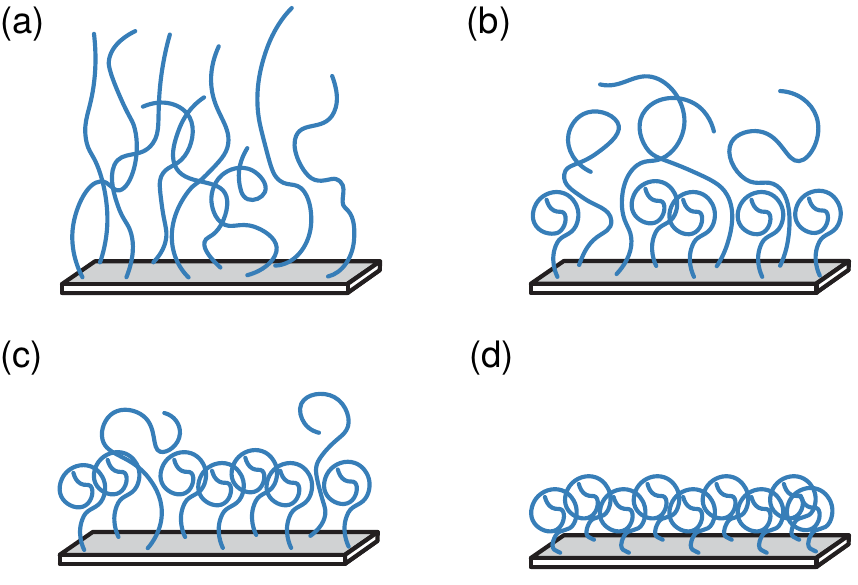}
\caption{(a) A fully swollen brush, (b) a coexisting brush with overlapping proteins in the outer layer, (c) a coexisting brush with isolated proteins in the outer layer, and (d) a fully condensed brush.}
\label{fig:trans}
\end{figure}

The features of the distributions of Block 3 and Block 5 illustrate that the coexisting brushes consist of two populations of chains, collapsed and stretched, as shown by the schematic in Fig.~\ref{fig:blocki_schem}. 
The inner condensed layer is comprised of all the collapsed chains and the beginning portion of the stretched chains, whereas the outer diffuse layer includes only the remaining portion of the stretched chains. As $I$ increases, the population of stretched chains transfers to that of the collapsed conformation. Only the stretched conformation is sensitive to the ionic strength, which becomes less extended as screening strength increases. This underlying picture of the coexisting morphology in NFH brushes is consistent with previous theoretical and experimental results observed in synthetic polyelectrolyte brushes \cite{klushin_microphase_2001, ballauff_phase_2016}.

The insight obtained by studying the morphology of coexisting brushes helps to explain the dramatic height change in response to ionic strength, as observed in Fig.~\ref{fig:prof} and in the experiments of the Kumar group \cite{srinivasan_stimuli-sensitive_2014}.
At very low ionic strengths ($I < 4$ mM), all NFH proteins within the brushes take the stretched conformation, as shown in Fig.~\ref{fig:trans_a}. At the critical ionic strength $I = 4$ mM, the screened electrostatic repulsion is comparable to the hydrophobic attraction such that individual proteins have a noticeable probability to transfer to the collapsed conformation. 
With increasing $I$, the number of proteins in the collapsed conformation increases, leading to the formation of the condensed inner layer of the coexisting brush. 
The outer layer of the coexisting brush can be viewed as being "grafted" upon the inner layer with effective grafting density $\sigma_{out}$. Particularly, at $I = 4$ mM, $\sigma_{out}\approx\sigma$ because almost all the proteins remain in the stretched conformation. For 4 mM $< I < 6$ mM, $\sigma_{out}$ drops dramatically as a significant number of stretched proteins collapse into the inner layer. 
However, as shown in Fig.~\ref{fig:trans_b}, $\sigma_{out}$ is still large enough such that proteins in the outer layer remain overlapped with each other. According to the Alexander-de Gennes model for the overlapping brush, the brush height is proportional to the effective grafting density, i.e. $H \sim \sigma_{out}$, which explains the dramatic height decrease in this regime \cite{borisov_diagram_1994}. For $I > 6$ mM, $\sigma_{out}$ continues decreasing such that the distance between stretched proteins is larger than the pervaded size of individual proteins in the lateral direction (i.e., $\sim\sigma_{out}^{-1/2}> R_{g,xy}$). 
As shown in Fig.~\ref{fig:trans_c}, proteins in the outer layer behave isolated in this regime, and the brush height is thus insensitive to the further decrease of grafting density, i.e., $H \sim \sigma_{out}^0$. The slow decrease of $H$ as $I$ increases in this regime is solely attributed to the screening effect on the electrostatic repulsion of the stretched proteins. 
At very high ionic strengths $I > 50$ mM, the electrostatic repulsion is highly screened, which leads to the collapse of all the proteins and the eventual disappearance of the outer diffuse layer (see Fig.~\ref{fig:trans_d}). 
Therefore, the dramatic height change characterizes the transition from the overlapping state to the isolated state in the outer layer of the coexisting brush induced by the electrostatic screening effect.      

\subsection{Characterization of Microstructure from Scattering and Force Spectra}

\begin{figure}[hbt!]
    \centering
    \includegraphics[width=0.90\linewidth]{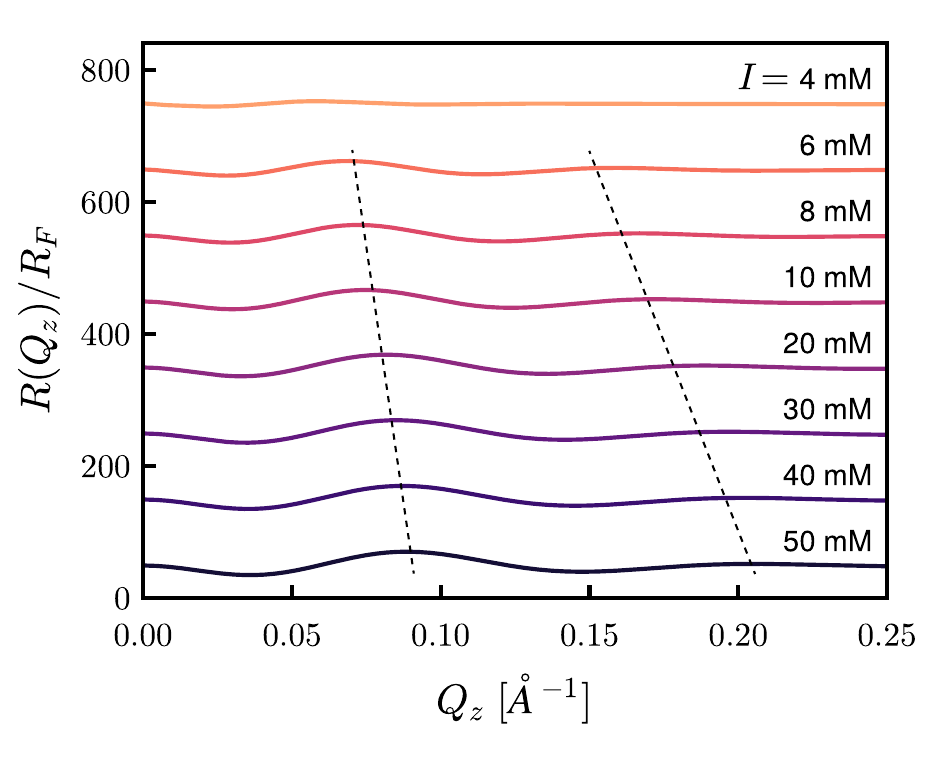} 
    \caption{Reflectivity spectra of NFH brushes at different ionic strengths calculated using protein density profiles. The reflected intensity $R$ (normalized by the Fresnel reflectivity $R_F$) is plotted against the specular part of the momentum transfer vector $Q_z$. For clarity, the curves are offset vertically by 100 as $I$ increases. The dashed guidelines illustrate the increase in periodicity of the oscillations.}
    \label{fig:refs}
\end{figure}

As shown in the above subsection, NFH brushes exhibit non-trivial morphological behaviors which cannot be fully captured by the overall brush height obtained from common characterization tools such as AFM and spectroscopic ellipsometry. Reflectivity is a scattering technique which detects the detailed structure of buried layered interfaces by utilizing the interference of X-rays or neutrons in the sample. Like all other scattering techniques, it is challenging to obtain density profiles directly from reflectivity spectra because the scattering intensity is measured in reciprocal space \cite{skoda_recent_2019, xu_formation_2013}. The intensity $R$ of a reflected X-ray beam as a function of the specular part of the momentum transfer vector $Q_z$ and the gradient of the electron density distribution $d\rho_e(z)/dz$ is given by:
\begin{equation}\label{eq:master}
R(Q_z) = R_F \left| \frac{1}{\rho_e^{\infty}}\int \textrm{dz}\ \frac{d\rho_e(z)}{dz} \exp{(iQ_z z)} \right| ^2,
\end{equation}
where $R_F$ represents the Fresnel reflectivity of an ideal, sharp surface and $\rho_e^{\infty}$ is the electron density of the bulk media far away from the substrate \cite{russell_x-ray_1990}. Here, we take $\rho_e^\infty = \rho_{e,s} = 0.33\ e\ \mathrm{\angstrom}^{-3}$ as the value of water. The local electron density $\rho_e(z)$ can be calculated from the distribution of each species $\alpha$ based on the following linear combination: $\rho_e(z) = \sum_{\alpha}\rho_{e,\alpha}\phi_\alpha(z)$. $\alpha$ includes the protein, solvent, and \ch{SiO2} substrate with the corresponding electron densities $\rho_{e,\alpha}$ as 0.95 $e\ \mathrm{\angstrom}^{-3}$, 0.33 $e\ \mathrm{\angstrom}^{-3}$, and 2.32 $e\ \mathrm{\angstrom}^{-3}$, respectively \cite{braslau_surface_1985}. 

The density profiles obtained from our theory can be used to investigate the evolution of the scattering intensity with changing ionic strength. 
Fig.~\ref{fig:refs} plots the reflectivity (normalized by the Fresnel reflectivity $R_F$) for different $I$. At $I = 4$ mM, $R(Q_z)$ is nearly featureless, indicating the absence of any condensed layer. For $I \geq 6$ mM, $R(Q_z)$ exhibits an oscillatory signature, which signifies the existence of a condensed layer. 
In this regime, it can be clearly seen from Fig.~\ref{fig:refs} that the ionic strength affects both the periodicity and amplitude of the oscillations.  The increase in periodicity (denoted by $\Delta Q_z$) with $I$ indicates that the condensed layer becomes thinner since the layer thickness $d$ is characterized by the inverse of the periodicity (i.e., $d=2\pi / \Delta Q_z$). On the other hand, the increase in the amplitude of the oscillations signifies condensed layers with sharper interfaces. 
Using reflectivity spectroscopy, similar oscillatory signatures have been observed in neutral polymer brushes in the presence of a condensed layer \cite{reinhardt_fine-tuning_2013, laloyaux_surface_2010, yim_evidence_2005}.
Our results demonstrate the effectiveness of reflectivity spectroscopy in detecting the morphological change of protein brushes in terms of the existence and microstructure of the condensed layer.        

\begin{figure}[hbt!]
    \centering
        \includegraphics[width=0.90\linewidth]{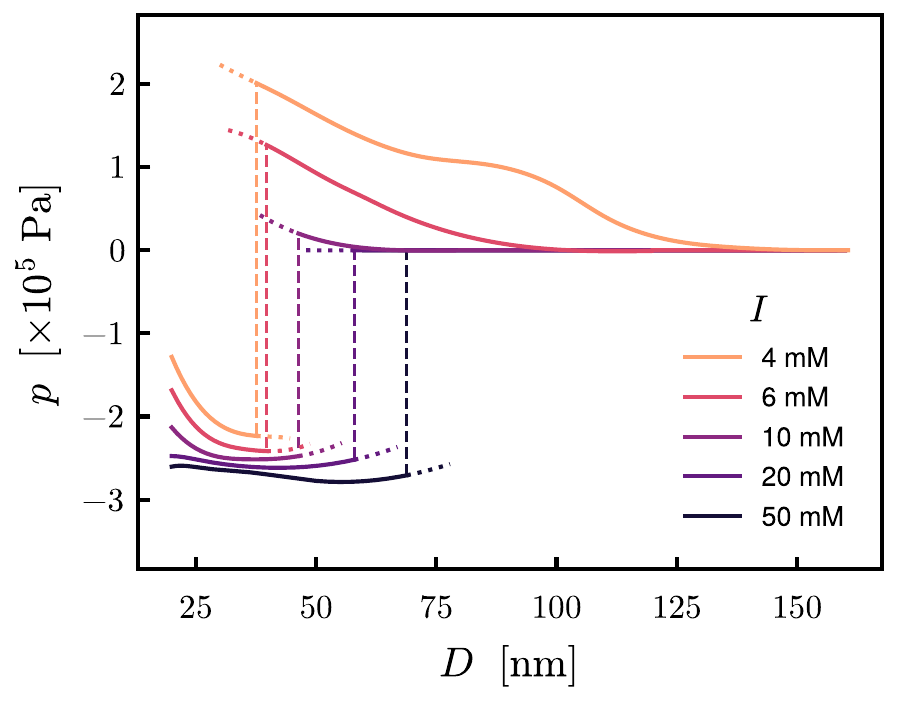} 
    \caption{Force spectra between two opposing substrates grafted with NFH brushes. The force per unit area $p$ is plotted against the separation distance $D$ at different ionic strengths. Vertical dashed lines locate the discontinuous transition of two separated condensed layer from both brushes to a single, merged condensed layer. The dotted lines denote the metastable region beyond the transition point.}
    \label{fig:fprof}
\end{figure}

Another approach to obtain the brush microstructure is through characterizing its mechanical response using force spectroscopy techniques such as the surface force apparatus (SFA) \cite{balastre_study_2002, drobek_compressing_2005}. 
By tracking the free energy and the corresponding brush morphology as a function of the separation distance, our theory enables the investigation of the force spectra between two opposing substrates grafted by NFH brushes. The force per unit area is quantified by the disjoining pressure $p$ at a given separation $D$ between the two substrates. It is calculated from the derivative of the free energy $F$ as $p = - (\partial F/\partial D)_{\lambda_\pm} - p_b$, where $p_b = c_+^b + c_-^b$ is the bulk osmotic pressure from the connected reservoir. Fig.~\ref{fig:fprof} shows the force spectra for different ionic strengths. For all $I$, the force is repulsive (i.e. $p > 0$) at large separations as a result of the dominant electrostatic repulsion in this region. 
It is interesting to note that a signature of a shoulder appears at $75 \textrm{ nm} < D < 100$ nm for $I =$ 4 mM, which is unique compared to the spectra at other ionic strengths. This signifies a morphological change as two swollen brushes approach each other. As $D$ decreases, the brushes begin to overlap, which enhances the electrostatic repulsion and induces the collapse of proteins to reduce the local charge density. In this regime, the condensed proteins are sparsely distributed and each protein collapses individually. The force is thus nearly constant as reflected by the shoulder in the spectra. 
The ``shoulder'' signature in the force spectra can be illuminated by the end-block density distribution as shown in Fig. 8.  For $75 < D < 100$ nm, the plateau in the region of 15 nm $< z < 30$ nm is almost maintained as $D$ decreases. There is a slight transfer of the density from the outer edge of the swollen layer to the inner condensed layer, indicating the independent collapse of individual proteins.
This signature is absent for brushes at $I \geq$ 6 mM, as the condensed layers are dense enough such that subsequent collapses of proteins cause progressively stronger energy penalties. Therefore, while reflectivity is sensitive to the condensed morphology, force spectroscopy is an effective tool for detecting the swollen morphology.

\begin{figure}
    \centering
    \includegraphics[width=0.90\linewidth]{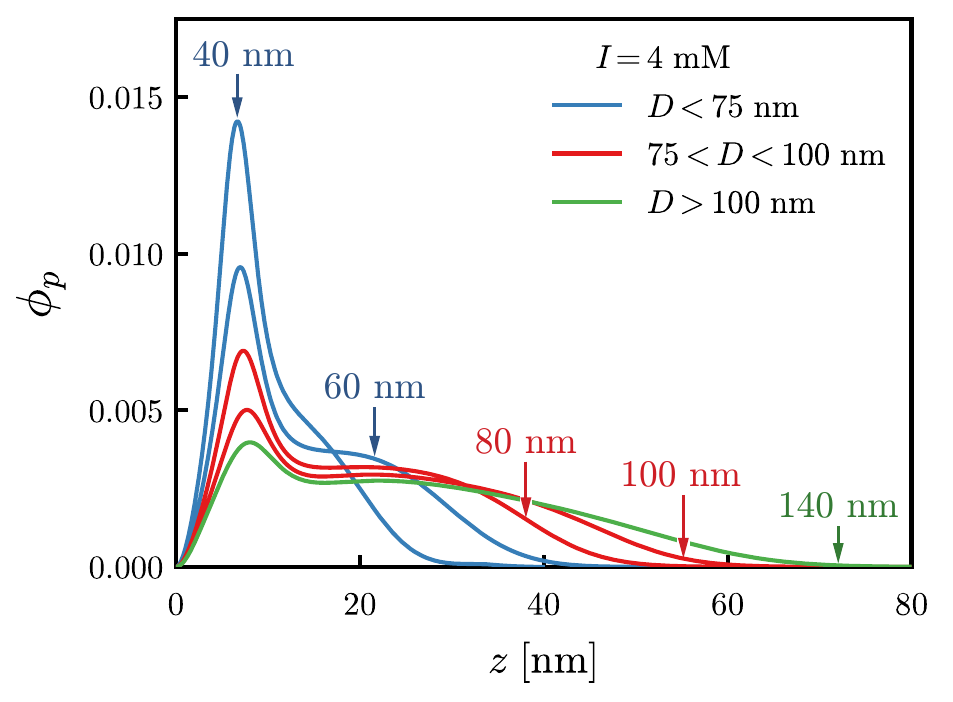}
    \caption{Representative end-block distributions of one of the two opposing brushes at $I=4$ mM. Distributions are labelled with their corresponding separation distances $D$. The regime $75 < D < 100$ nm corresponds to the shoulder signature in the force spectra in Fig.~\ref{fig:fprof}.}
    \label{fig:fdens}
\end{figure}

As shown in Fig.~\ref{fig:fprof}, $p$ drops discontinuously to a negative value at short separations for all $I$. This is caused by the competition between the electrostatic repulsion dominant at large $D$ and the hydrophobic attraction between opposing brushes which dominates at small $D$. 
At a critical $D$, the two separated condensed layers from both brushes merge into a single condensed layer that occupies the entire space between the two substrates. As $I$ decreases, the electrostatic repulsion is less screened, which leads to a shift of the transition to smaller $D$ and a deeper attractive well. It is worth noting that the brush morphology and interaction is significantly affected by the dielectric inhomogeneity of the system, particularly at small separation distances. Furthermore, at very small separations, the excluded volume effect as a result of incompressibility and chain packing will play a significant role, where the pressure will rapidly increase and eventually become positive. Our work thus elucidates the non-trivial mechanical response and the corresponding morphological change between interacting brushes. 
\section{Conclusions}
In this work, we have applied a continuous-space self-consistent field theory to study the morphological response of a neurofilament-derived protein brush to ionic strength. A coarse-graining approach based on a multi-block charged macromolecular model has been developed to capture the chemical identity of the amino acid sequence. 
For varying ionic strengths, the height of NFH brushes at pH 2.4 predicted by our theory is in good agreement with the experimental data reported in the literature. NFH brushes exhibit three distinct morphological regimes: swollen brushes at low $I$, condensed brushes at high $I$, and coexisting brushes at intermediate $I$ which contain a dense inner layer and a diffuse outer layer. 
Our theory enables the study of brush microstructure in terms of density distributions of constituent residues. We find that the dramatic height change observed in experiments originates from the transition between the overlapping state and the isolated state in the outer layer of the coexisting brushes induced by electrostatic screening. 
The evolution of the scattering behavior and mechanical property accompanying the morphological change has also been investigated. The appearance of the oscillatory signature in reflectivity spectra characterizes the existence of the condensed inner layer. Both the periodicity and the amplitude of the oscillations increase as $I$ increases, signifying thinner condensed layers with sharper interfaces. Furthermore, the force spectra between two opposing swollen brushes shows a signature of a shoulder due to the independent collapses of individual proteins. 
Our results demonstrate that reflectivity spectroscopy is sensitive to brushes with condensed layers and force spectroscopy is an effective tool for detecting the microstructure of brushes with swollen morphology.  

Although the current work focuses on NFH brushes at pH 2.4, our theory can be straightforwardly generalized to brushes with any amino acid sequence, which facilitates the study of the effect of genetic and chemical modifications, such as phosphorylation. Our theory can also incorporate an explicit treatment of local proton concentration. The current work focuses on the comparison with experimental data measured at low pH (pH = 2.4). Under this condition of high proton concentration, the difference between the local and bulk proton concentrations can be neglected. However, at high pH, the local proton concentration can be significantly different from the bulk value, which necessitates an explicit treatment of local proton exchange \cite{balzer_electroresponse_2023}. This treatment is important to accurately capture the response of brushes to external pH stimuli, particularly for proteins consisting of weak acid/base residues. We can also apply the theory to model brushes comprised of a mixture of different proteins to better represent neurofilaments. Lastly, the calculation can be performed at high dimensions to investigate brush patterning parallel to the substrate and interactions between two cylindrical brushes. 
Our theory provides a high-throughput computational platform which bridges the chemical structure of protein brushes and their morphological, scattering, and mechanical responses. This is important for the fundamental understanding of neurofilaments in axonal physiology, which plays a key role in the rational design of both stimuli-sensitive biomaterials and therapies for neurodegenerative diseases.

\begin{acknowledgement}
This material is based upon work supported by the National Science Foundation Graduate Research Fellowship under Grant No. DGE 2146752 to TJY and EAD. This research used the computational resource provided by the Kenneth S. Pitzer Center for Theoretical Chemistry. The work was also supported by NIH R01GM122375 to SK.
\end{acknowledgement}

\begin{suppinfo}
NFH sequence and model parameters for amino acids; Derivation of self-consistent field theory for protein brushes; Sensitivity analysis of the coarse-graining procedure; Ionic density profiles near the substrate.
\end{suppinfo}

\bibliography{nfh_20230828}

\providecommand{\latin}[1]{#1}
\makeatletter
\providecommand{\doi}
  {\begingroup\let\do\@makeother\dospecials
  \catcode`\{=1 \catcode`\}=2 \doi@aux}
\providecommand{\doi@aux}[1]{\endgroup\texttt{#1}}
\makeatother
\providecommand*\mcitethebibliography{\thebibliography}
\csname @ifundefined\endcsname{endmcitethebibliography}  {\let\endmcitethebibliography\endthebibliography}{}
\begin{mcitethebibliography}{70}
\providecommand*\natexlab[1]{#1}
\providecommand*\mciteSetBstSublistMode[1]{}
\providecommand*\mciteSetBstMaxWidthForm[2]{}
\providecommand*\mciteBstWouldAddEndPuncttrue
  {\def\EndOfBibitem{\unskip.}}
\providecommand*\mciteBstWouldAddEndPunctfalse
  {\let\EndOfBibitem\relax}
\providecommand*\mciteSetBstMidEndSepPunct[3]{}
\providecommand*\mciteSetBstSublistLabelBeginEnd[3]{}
\providecommand*\EndOfBibitem{}
\mciteSetBstSublistMode{f}
\mciteSetBstMaxWidthForm{subitem}{(\alph{mcitesubitemcount})}
\mciteSetBstSublistLabelBeginEnd
  {\mcitemaxwidthsubitemform\space}
  {\relax}
  {\relax}

\bibitem[Yuan \latin{et~al.}(2017)Yuan, Rao, Veeranna, and Nixon]{yuan_neurofilaments_2017}
Yuan,~A.; Rao,~M.~V.; Veeranna; Nixon,~R.~A. Neurofilaments and {Neurofilament} {Proteins} in {Health} and {Disease}. \emph{Cold Spring Harbor Perspect. Biol.} \textbf{2017}, \emph{9}, a018309\relax
\mciteBstWouldAddEndPuncttrue
\mciteSetBstMidEndSepPunct{\mcitedefaultmidpunct}
{\mcitedefaultendpunct}{\mcitedefaultseppunct}\relax
\EndOfBibitem
\bibitem[Zhu \latin{et~al.}(1997)Zhu, Couillard-Després, and Julien]{zhu_delayed_1997}
Zhu,~Q.; Couillard-Després,~S.; Julien,~J.-P. Delayed {Maturation} of {Regenerating} {Myelinated} {Axons} in {Mice} {Lacking} {Neurofilaments}. \emph{Exp. Neurol.} \textbf{1997}, \emph{148}, 299--316\relax
\mciteBstWouldAddEndPuncttrue
\mciteSetBstMidEndSepPunct{\mcitedefaultmidpunct}
{\mcitedefaultendpunct}{\mcitedefaultseppunct}\relax
\EndOfBibitem
\bibitem[Yuan \latin{et~al.}(2012)Yuan, Rao, ~, and Nixon]{yuan_neurofilaments_2012}
Yuan,~A.; Rao,~M.~V.; ~,~V.; Nixon,~R.~A. Neurofilaments at a glance. \emph{J. Cell Sci.} \textbf{2012}, \emph{125}, 3257--3263\relax
\mciteBstWouldAddEndPuncttrue
\mciteSetBstMidEndSepPunct{\mcitedefaultmidpunct}
{\mcitedefaultendpunct}{\mcitedefaultseppunct}\relax
\EndOfBibitem
\bibitem[Barro \latin{et~al.}(2020)Barro, Chitnis, and Weiner]{barro_blood_2020}
Barro,~C.; Chitnis,~T.; Weiner,~H.~L. Blood neurofilament light: a critical review of its application to neurologic disease. \emph{Ann. Clin. Transl. Neurol.} \textbf{2020}, \emph{7}, 2508--2523\relax
\mciteBstWouldAddEndPuncttrue
\mciteSetBstMidEndSepPunct{\mcitedefaultmidpunct}
{\mcitedefaultendpunct}{\mcitedefaultseppunct}\relax
\EndOfBibitem
\bibitem[Lavedan \latin{et~al.}(2002)Lavedan, Buchholtz, Nussbaum, Albin, and Polymeropoulos]{lavedan_mutation_2002}
Lavedan,~C.; Buchholtz,~S.; Nussbaum,~R.~L.; Albin,~R.~L.; Polymeropoulos,~M.~H. A mutation in the human neurofilament {M} gene in {Parkinson}'s disease that suggests a role for the cytoskeleton in neuronal degeneration. \emph{Neurosci. Lett.} \textbf{2002}, \emph{322}, 57--61\relax
\mciteBstWouldAddEndPuncttrue
\mciteSetBstMidEndSepPunct{\mcitedefaultmidpunct}
{\mcitedefaultendpunct}{\mcitedefaultseppunct}\relax
\EndOfBibitem
\bibitem[Langer and Tirrell(2004)Langer, and Tirrell]{langer_designing_2004}
Langer,~R.; Tirrell,~D.~A. Designing materials for biology and medicine. \emph{Nature} \textbf{2004}, \emph{428}, 487--492\relax
\mciteBstWouldAddEndPuncttrue
\mciteSetBstMidEndSepPunct{\mcitedefaultmidpunct}
{\mcitedefaultendpunct}{\mcitedefaultseppunct}\relax
\EndOfBibitem
\bibitem[Pan \latin{et~al.}(2020)Pan, Aaron~Lau, Messersmith, Lu, and Zhao]{pan_interfacial_2020}
Pan,~F.; Aaron~Lau,~K.~H.; Messersmith,~P.~B.; Lu,~J.~R.; Zhao,~X. Interfacial {Assembly} {Inspired} by {Marine} {Mussels} and {Antifouling} {Effects} of {Polypeptoids}: {A} {Neutron} {Reflection} {Study}. \emph{Langmuir} \textbf{2020}, \emph{36}, 12309--12318\relax
\mciteBstWouldAddEndPuncttrue
\mciteSetBstMidEndSepPunct{\mcitedefaultmidpunct}
{\mcitedefaultendpunct}{\mcitedefaultseppunct}\relax
\EndOfBibitem
\bibitem[Das \latin{et~al.}(2015)Das, Banik, Chen, Sinha, and Mukherjee]{das_polyelectrolyte_2015}
Das,~S.; Banik,~M.; Chen,~G.; Sinha,~S.; Mukherjee,~R. Polyelectrolyte brushes: theory, modelling, synthesis and applications. \emph{Soft Matter} \textbf{2015}, \emph{11}, 8550--8583\relax
\mciteBstWouldAddEndPuncttrue
\mciteSetBstMidEndSepPunct{\mcitedefaultmidpunct}
{\mcitedefaultendpunct}{\mcitedefaultseppunct}\relax
\EndOfBibitem
\bibitem[Conrad and Robertson(2019)Conrad, and Robertson]{conrad_towards_2019}
Conrad,~J.~C.; Robertson,~M.~L. Towards mimicking biological function with responsive surface-grafted polymer brushes. \emph{Curr. Opin. Solid State Mater. Sci.} \textbf{2019}, \emph{23}, 1--12\relax
\mciteBstWouldAddEndPuncttrue
\mciteSetBstMidEndSepPunct{\mcitedefaultmidpunct}
{\mcitedefaultendpunct}{\mcitedefaultseppunct}\relax
\EndOfBibitem
\bibitem[Blum \latin{et~al.}(2022)Blum, Yin, Lin, Oliver, Kammeyer, Thompson, Gilson, and Gianneschi]{blum_stimuli_2022}
Blum,~A.~P.; Yin,~J.; Lin,~H.~H.; Oliver,~B.~A.; Kammeyer,~J.~K.; Thompson,~M.~P.; Gilson,~M.~K.; Gianneschi,~N.~C. Stimuli {Induced} {Uptake} of {Protein}-{Like} {Peptide} {Brush} {Polymers}. \emph{Chem. - Eur. J.} \textbf{2022}, \emph{28}, e202103438\relax
\mciteBstWouldAddEndPuncttrue
\mciteSetBstMidEndSepPunct{\mcitedefaultmidpunct}
{\mcitedefaultendpunct}{\mcitedefaultseppunct}\relax
\EndOfBibitem
\bibitem[Wiarachai \latin{et~al.}(2016)Wiarachai, Vilaivan, Iwasaki, and Hoven]{wiarachai_clickable_2016}
Wiarachai,~O.; Vilaivan,~T.; Iwasaki,~Y.; Hoven,~V.~P. Clickable and {Antifouling} {Platform} of {Poly}[(propargyl methacrylate)-ran-(2-methacryloyloxyethyl phosphorylcholine)] for {Biosensing} {Applications}. \emph{Langmuir} \textbf{2016}, \emph{32}, 1184--1194\relax
\mciteBstWouldAddEndPuncttrue
\mciteSetBstMidEndSepPunct{\mcitedefaultmidpunct}
{\mcitedefaultendpunct}{\mcitedefaultseppunct}\relax
\EndOfBibitem
\bibitem[Xie \latin{et~al.}(2015)Xie, Tian, Wen, Xiao, Zhang, Liu, Hou, Li, Tian, and Jiang]{xie_chiral_2015}
Xie,~G.; Tian,~W.; Wen,~L.; Xiao,~K.; Zhang,~Z.; Liu,~Q.; Hou,~G.; Li,~P.; Tian,~Y.; Jiang,~L. Chiral recognition of l -tryptophan with beta-cyclodextrin-modified biomimetic single nanochannel. \emph{Chem. Commun.} \textbf{2015}, \emph{51}, 3135--3138\relax
\mciteBstWouldAddEndPuncttrue
\mciteSetBstMidEndSepPunct{\mcitedefaultmidpunct}
{\mcitedefaultendpunct}{\mcitedefaultseppunct}\relax
\EndOfBibitem
\bibitem[Welch \latin{et~al.}(2011)Welch, Rastogi, and Ober]{welch_polymer_2011}
Welch,~M.; Rastogi,~A.; Ober,~C. Polymer brushes for electrochemical biosensors. \emph{Soft Matter} \textbf{2011}, \emph{7}, 297--302\relax
\mciteBstWouldAddEndPuncttrue
\mciteSetBstMidEndSepPunct{\mcitedefaultmidpunct}
{\mcitedefaultendpunct}{\mcitedefaultseppunct}\relax
\EndOfBibitem
\bibitem[Ito and Soon~Park(2000)Ito, and Soon~Park]{ito_signal-responsive_2000}
Ito,~Y.; Soon~Park,~Y. Signal-responsive gating of porous membranes by polymer brushes. \emph{Polym. Adv. Technol.} \textbf{2000}, \emph{11}, 136--144\relax
\mciteBstWouldAddEndPuncttrue
\mciteSetBstMidEndSepPunct{\mcitedefaultmidpunct}
{\mcitedefaultendpunct}{\mcitedefaultseppunct}\relax
\EndOfBibitem
\bibitem[Yameen \latin{et~al.}(2009)Yameen, Ali, Neumann, Ensinger, Knoll, and Azzaroni]{yameen_single_2009}
Yameen,~B.; Ali,~M.; Neumann,~R.; Ensinger,~W.; Knoll,~W.; Azzaroni,~O. Single {Conical} {Nanopores} {Displaying} {pH}-{Tunable} {Rectifying} {Characteristics}. {Manipulating} {Ionic} {Transport} {With} {Zwitterionic} {Polymer} {Brushes}. \emph{J. Am. Chem. Soc.} \textbf{2009}, \emph{131}, 2070--2071\relax
\mciteBstWouldAddEndPuncttrue
\mciteSetBstMidEndSepPunct{\mcitedefaultmidpunct}
{\mcitedefaultendpunct}{\mcitedefaultseppunct}\relax
\EndOfBibitem
\bibitem[Kelby \latin{et~al.}(2011)Kelby, Wang, and Huck]{kelby_controlled_2011}
Kelby,~T.~S.; Wang,~M.; Huck,~W.~T. Controlled {Folding} of {2D} {Au}–{Polymer} {Brush} {Composites} into {3D} {Microstructures}. \emph{Adv. Funct. Mater.} \textbf{2011}, \emph{21}, 652--657\relax
\mciteBstWouldAddEndPuncttrue
\mciteSetBstMidEndSepPunct{\mcitedefaultmidpunct}
{\mcitedefaultendpunct}{\mcitedefaultseppunct}\relax
\EndOfBibitem
\bibitem[Ionov \latin{et~al.}(2006)Ionov, Stamm, and Diez]{ionov_reversible_2006}
Ionov,~L.; Stamm,~M.; Diez,~S. Reversible {Switching} of {Microtubule} {Motility} {Using} {Thermoresponsive} {Polymer} {Surfaces}. \emph{Nano Lett.} \textbf{2006}, \emph{6}, 1982--1987\relax
\mciteBstWouldAddEndPuncttrue
\mciteSetBstMidEndSepPunct{\mcitedefaultmidpunct}
{\mcitedefaultendpunct}{\mcitedefaultseppunct}\relax
\EndOfBibitem
\bibitem[Kobayashi and Takahara(2010)Kobayashi, and Takahara]{kobayashi_tribological_2010}
Kobayashi,~M.; Takahara,~A. Tribological properties of hydrophilic polymer brushes under wet conditions. \emph{Chem. Rec.} \textbf{2010}, \emph{10}, 208--216\relax
\mciteBstWouldAddEndPuncttrue
\mciteSetBstMidEndSepPunct{\mcitedefaultmidpunct}
{\mcitedefaultendpunct}{\mcitedefaultseppunct}\relax
\EndOfBibitem
\bibitem[Liechty \latin{et~al.}(2010)Liechty, Kryscio, Slaughter, and Peppas]{liechty_polymers_2010}
Liechty,~W.~B.; Kryscio,~D.~R.; Slaughter,~B.~V.; Peppas,~N.~A. Polymers for {Drug} {Delivery} {Systems}. \emph{Annu. Rev. Chem. Biomol. Eng.} \textbf{2010}, \emph{1}, 149--173\relax
\mciteBstWouldAddEndPuncttrue
\mciteSetBstMidEndSepPunct{\mcitedefaultmidpunct}
{\mcitedefaultendpunct}{\mcitedefaultseppunct}\relax
\EndOfBibitem
\bibitem[Srinivasan \latin{et~al.}(2014)Srinivasan, Bhagawati, Ananthanarayanan, and Kumar]{srinivasan_stimuli-sensitive_2014}
Srinivasan,~N.; Bhagawati,~M.; Ananthanarayanan,~B.; Kumar,~S. Stimuli-sensitive intrinsically disordered protein brushes. \emph{Nat. Commun.} \textbf{2014}, \emph{5}, 5145\relax
\mciteBstWouldAddEndPuncttrue
\mciteSetBstMidEndSepPunct{\mcitedefaultmidpunct}
{\mcitedefaultendpunct}{\mcitedefaultseppunct}\relax
\EndOfBibitem
\bibitem[Israels \latin{et~al.}(1994)Israels, Leermakers, Fleer, and Zhulina]{israels_charged_1994}
Israels,~R.; Leermakers,~F. A.~M.; Fleer,~G.~J.; Zhulina,~E.~B. Charged {Polymeric} {Brushes}: {Structure} and {Scaling} {Relations}. \emph{Macromolecules} \textbf{1994}, \emph{27}, 3249--3261\relax
\mciteBstWouldAddEndPuncttrue
\mciteSetBstMidEndSepPunct{\mcitedefaultmidpunct}
{\mcitedefaultendpunct}{\mcitedefaultseppunct}\relax
\EndOfBibitem
\bibitem[Chen \latin{et~al.}(2017)Chen, Cordero, Tran, and Ober]{chen_50th_2017}
Chen,~W.-L.; Cordero,~R.; Tran,~H.; Ober,~C.~K. 50th {Anniversary} {Perspective}: {Polymer} {Brushes}: {Novel} {Surfaces} for {Future} {Materials}. \emph{Macromolecules} \textbf{2017}, \emph{50}, 4089--4113\relax
\mciteBstWouldAddEndPuncttrue
\mciteSetBstMidEndSepPunct{\mcitedefaultmidpunct}
{\mcitedefaultendpunct}{\mcitedefaultseppunct}\relax
\EndOfBibitem
\bibitem[Fuchs and Cleveland(1998)Fuchs, and Cleveland]{fuchs_structural_1998}
Fuchs,~E.; Cleveland,~D.~W. A {Structural} {Scaffolding} of {Intermediate} {Filaments} in {Health} and {Disease}. \emph{Science} \textbf{1998}, \emph{279}, 514--519\relax
\mciteBstWouldAddEndPuncttrue
\mciteSetBstMidEndSepPunct{\mcitedefaultmidpunct}
{\mcitedefaultendpunct}{\mcitedefaultseppunct}\relax
\EndOfBibitem
\bibitem[Zhulina and Leermakers(2010)Zhulina, and Leermakers]{zhulina_polymer_2010}
Zhulina,~E.; Leermakers,~F. The {Polymer} {Brush} {Model} of {Neurofilament} {Projections}: {Effect} of {Protein} {Composition}. \emph{Biophys. J.} \textbf{2010}, \emph{98}, 462--469\relax
\mciteBstWouldAddEndPuncttrue
\mciteSetBstMidEndSepPunct{\mcitedefaultmidpunct}
{\mcitedefaultendpunct}{\mcitedefaultseppunct}\relax
\EndOfBibitem
\bibitem[Elder \latin{et~al.}(1998)Elder, Friedrich, Kang, Bosco, Gourov, Tu, Zhang, Lee, and Lazzarini]{elder_requirement_1998}
Elder,~G.~A.; Friedrich,~V.~L.,~Jr.; Kang,~C.; Bosco,~P.; Gourov,~A.; Tu,~P.-H.; Zhang,~B.; Lee,~V. M.-Y.; Lazzarini,~R.~A. Requirement of {Heavy} {Neurofilament} {Subunit} in the {Development} of {Axons} with {Large} {Calibers}. \emph{J. Cell Biol.} \textbf{1998}, \emph{143}, 195--205\relax
\mciteBstWouldAddEndPuncttrue
\mciteSetBstMidEndSepPunct{\mcitedefaultmidpunct}
{\mcitedefaultendpunct}{\mcitedefaultseppunct}\relax
\EndOfBibitem
\bibitem[Zhu \latin{et~al.}(1998)Zhu, Lindenbaum, Levavasseur, Jacomy, and Julien]{zhu_disruption_1998}
Zhu,~Q.; Lindenbaum,~M.; Levavasseur,~F.; Jacomy,~H.; Julien,~J.-P. Disruption of the {NF}-{H} {Gene} {Increases} {Axonal} {Microtubule} {Content} and {Velocity} of {Neurofilament} {Transport}: {Relief} of {Axonopathy} {Resulting} from the {Toxin} Beta,Beta'-{Iminodipropionitrile}. \emph{J. Cell Biol.} \textbf{1998}, \emph{143}, 183--193\relax
\mciteBstWouldAddEndPuncttrue
\mciteSetBstMidEndSepPunct{\mcitedefaultmidpunct}
{\mcitedefaultendpunct}{\mcitedefaultseppunct}\relax
\EndOfBibitem
\bibitem[Rao \latin{et~al.}(1998)Rao, Houseweart, Williamson, Crawford, Folmer, and Cleveland]{rao_neurofilament-dependent_1998}
Rao,~M.~V.; Houseweart,~M.~K.; Williamson,~T.~L.; Crawford,~T.~O.; Folmer,~J.; Cleveland,~D.~W. Neurofilament-dependent {Radial} {Growth} of {Motor} {Axons} and {Axonal} {Organization} of {Neurofilaments} {Does} {Not} {Require} the {Neurofilament} {Heavy} {Subunit} ({NF}-{H}) or {Its} {Phosphorylation}. \emph{J. Cell Biol.} \textbf{1998}, \emph{143}, 171--181\relax
\mciteBstWouldAddEndPuncttrue
\mciteSetBstMidEndSepPunct{\mcitedefaultmidpunct}
{\mcitedefaultendpunct}{\mcitedefaultseppunct}\relax
\EndOfBibitem
\bibitem[Bhagawati \latin{et~al.}(2016)Bhagawati, Rubashkin, Lee, Ananthanarayanan, Weaver, and Kumar]{bhagawati_site-specific_2016}
Bhagawati,~M.; Rubashkin,~M.~G.; Lee,~J.~P.; Ananthanarayanan,~B.; Weaver,~V.~M.; Kumar,~S. Site-{Specific} {Modulation} of {Charge} {Controls} the {Structure} and {Stimulus} {Responsiveness} of {Intrinsically} {Disordered} {Peptide} {Brushes}. \emph{Langmuir} \textbf{2016}, \emph{32}, 5990--5996\relax
\mciteBstWouldAddEndPuncttrue
\mciteSetBstMidEndSepPunct{\mcitedefaultmidpunct}
{\mcitedefaultendpunct}{\mcitedefaultseppunct}\relax
\EndOfBibitem
\bibitem[Lei \latin{et~al.}(2018)Lei, Lee, Francis, and Kumar]{lei_structural_2018}
Lei,~R.; Lee,~J.~P.; Francis,~M.~B.; Kumar,~S. Structural {Regulation} of a {Neurofilament}-{Inspired} {Intrinsically} {Disordered} {Protein} {Brush} by {Multisite} {Phosphorylation}. \emph{Biochemistry} \textbf{2018}, \emph{57}, 4019--4028\relax
\mciteBstWouldAddEndPuncttrue
\mciteSetBstMidEndSepPunct{\mcitedefaultmidpunct}
{\mcitedefaultendpunct}{\mcitedefaultseppunct}\relax
\EndOfBibitem
\bibitem[Hest and A. Tirrell(2001)Hest, and A. Tirrell]{hest_protein-based_2001}
Hest,~J. C. M.~v.; A. Tirrell,~D. Protein-based materials, toward a new level of structural control. \emph{Chem. Commun.} \textbf{2001}, \emph{0}, 1897--1904\relax
\mciteBstWouldAddEndPuncttrue
\mciteSetBstMidEndSepPunct{\mcitedefaultmidpunct}
{\mcitedefaultendpunct}{\mcitedefaultseppunct}\relax
\EndOfBibitem
\bibitem[Alexander(1977)]{alexander_adsorption_1977}
Alexander,~S. Adsorption of chain molecules with a polar head a scaling description. \emph{J. Phys. (Paris)} \textbf{1977}, \emph{38}, 983--987\relax
\mciteBstWouldAddEndPuncttrue
\mciteSetBstMidEndSepPunct{\mcitedefaultmidpunct}
{\mcitedefaultendpunct}{\mcitedefaultseppunct}\relax
\EndOfBibitem
\bibitem[de~Gennes(1980)]{de_gennes_conformations_1980}
de~Gennes,~P.-G. Conformations of {Polymers} {Attached} to an {Interface}. \emph{Macromolecules} \textbf{1980}, \emph{13}, 1069--1075\relax
\mciteBstWouldAddEndPuncttrue
\mciteSetBstMidEndSepPunct{\mcitedefaultmidpunct}
{\mcitedefaultendpunct}{\mcitedefaultseppunct}\relax
\EndOfBibitem
\bibitem[Milner \latin{et~al.}(1988)Milner, Witten, and Cates]{milner_theory_1988}
Milner,~S.~T.; Witten,~T.~A.; Cates,~M.~E. Theory of the grafted polymer brush. \emph{Macromolecules} \textbf{1988}, \emph{21}, 2610--2619\relax
\mciteBstWouldAddEndPuncttrue
\mciteSetBstMidEndSepPunct{\mcitedefaultmidpunct}
{\mcitedefaultendpunct}{\mcitedefaultseppunct}\relax
\EndOfBibitem
\bibitem[{Milner, S. T.}(1991)]{milner_s_t_polymer_1991}
{Milner, S. T.} Polymer {Brushes}. \emph{Science} \textbf{1991}, \emph{251}, 905--914\relax
\mciteBstWouldAddEndPuncttrue
\mciteSetBstMidEndSepPunct{\mcitedefaultmidpunct}
{\mcitedefaultendpunct}{\mcitedefaultseppunct}\relax
\EndOfBibitem
\bibitem[Skvortsov \latin{et~al.}(1988)Skvortsov, Pavlushkov, Gorbunov, Zhulina, Borisov, and Pryamitsyn]{skvortsov_structure_1988}
Skvortsov,~A.~M.; Pavlushkov,~I.~V.; Gorbunov,~A.~A.; Zhulina,~Y.~B.; Borisov,~O.~V.; Pryamitsyn,~V.~A. Structure of densely grafted polymeric monolayers. \emph{Polym. Sci. U.S.S.R.} \textbf{1988}, \emph{30}, 1706--1715\relax
\mciteBstWouldAddEndPuncttrue
\mciteSetBstMidEndSepPunct{\mcitedefaultmidpunct}
{\mcitedefaultendpunct}{\mcitedefaultseppunct}\relax
\EndOfBibitem
\bibitem[Wijmans \latin{et~al.}(1992)Wijmans, Scheutjens, and Zhulina]{wijmans_self-consistent_1992}
Wijmans,~C.~M.; Scheutjens,~J. M. H.~M.; Zhulina,~E.~B. Self-consistent field theories for polymer brushes: lattice calculations and an asymptotic analytical description. \emph{Macromolecules} \textbf{1992}, \emph{25}, 2657--2665\relax
\mciteBstWouldAddEndPuncttrue
\mciteSetBstMidEndSepPunct{\mcitedefaultmidpunct}
{\mcitedefaultendpunct}{\mcitedefaultseppunct}\relax
\EndOfBibitem
\bibitem[Scheutjens and Fleer(1979)Scheutjens, and Fleer]{scheutjens_statistical_1979}
Scheutjens,~J. M. H.~M.; Fleer,~G.~J. Statistical theory of the adsorption of interacting chain molecules. 1. {Partition} function, segment density distribution, and adsorption isotherms. \emph{J. Phys. Chem.} \textbf{1979}, \emph{83}, 1619--1635\relax
\mciteBstWouldAddEndPuncttrue
\mciteSetBstMidEndSepPunct{\mcitedefaultmidpunct}
{\mcitedefaultendpunct}{\mcitedefaultseppunct}\relax
\EndOfBibitem
\bibitem[Scheutjens and Fleer(1980)Scheutjens, and Fleer]{scheutjens_statistical_1980}
Scheutjens,~J. M. H.~M.; Fleer,~G.~J. Statistical theory of the adsorption of interacting chain molecules. 2. {Train}, loop, and tail size distribution. \emph{J. Phys. Chem.} \textbf{1980}, \emph{84}, 178--190\relax
\mciteBstWouldAddEndPuncttrue
\mciteSetBstMidEndSepPunct{\mcitedefaultmidpunct}
{\mcitedefaultendpunct}{\mcitedefaultseppunct}\relax
\EndOfBibitem
\bibitem[Leermakers \latin{et~al.}(2010)Leermakers, Jho, and Zhulina]{leermakers_modeling_2010}
Leermakers,~F. A.~M.; Jho,~Y.-S.; Zhulina,~E.~B. Modeling of the {3RS} tau protein with self-consistent field method and {Monte} {Carlo} simulation. \emph{Soft Matter} \textbf{2010}, \emph{6}, 5533--5540\relax
\mciteBstWouldAddEndPuncttrue
\mciteSetBstMidEndSepPunct{\mcitedefaultmidpunct}
{\mcitedefaultendpunct}{\mcitedefaultseppunct}\relax
\EndOfBibitem
\bibitem[Zhulina and Leermakers(2007)Zhulina, and Leermakers]{zhulina_effect_2007}
Zhulina,~E.~B.; Leermakers,~F. A.~M. Effect of the {Ionic} {Strength} and {pH} on the {Equilibrium} {Structure} of a {Neurofilament} {Brush}. \emph{Biophys. J.} \textbf{2007}, \emph{93}, 1452--1463\relax
\mciteBstWouldAddEndPuncttrue
\mciteSetBstMidEndSepPunct{\mcitedefaultmidpunct}
{\mcitedefaultendpunct}{\mcitedefaultseppunct}\relax
\EndOfBibitem
\bibitem[Zhulina and Leermakers(2007)Zhulina, and Leermakers]{zhulina_self-consistent_2007}
Zhulina,~E.; Leermakers,~F. A {Self}-{Consistent} {Field} {Analysis} of the {Neurofilament} {Brush} with {Amino}-{Acid} {Resolution}. \emph{Biophys. J.} \textbf{2007}, \emph{93}, 1421--1430\relax
\mciteBstWouldAddEndPuncttrue
\mciteSetBstMidEndSepPunct{\mcitedefaultmidpunct}
{\mcitedefaultendpunct}{\mcitedefaultseppunct}\relax
\EndOfBibitem
\bibitem[Bianchi \latin{et~al.}(2020)Bianchi, Longhi, Grandori, and Brocca]{bianchi_relevance_2020}
Bianchi,~G.; Longhi,~S.; Grandori,~R.; Brocca,~S. Relevance of {Electrostatic} {Charges} in {Compactness}, {Aggregation}, and {Phase} {Separation} of {Intrinsically} {Disordered} {Proteins}. \emph{Int. J. Mol. Sci.} \textbf{2020}, \emph{21}, 6208\relax
\mciteBstWouldAddEndPuncttrue
\mciteSetBstMidEndSepPunct{\mcitedefaultmidpunct}
{\mcitedefaultendpunct}{\mcitedefaultseppunct}\relax
\EndOfBibitem
\bibitem[Lee \latin{et~al.}(2013)Lee, Kim, Chang, Jayanthi, and Gebremichael]{lee_effects_2013}
Lee,~J.; Kim,~S.; Chang,~R.; Jayanthi,~L.; Gebremichael,~Y. Effects of molecular model, ionic strength, divalent ions, and hydrophobic interaction on human neurofilament conformation. \emph{J. Chem. Phys.} \textbf{2013}, \emph{138}, 015103\relax
\mciteBstWouldAddEndPuncttrue
\mciteSetBstMidEndSepPunct{\mcitedefaultmidpunct}
{\mcitedefaultendpunct}{\mcitedefaultseppunct}\relax
\EndOfBibitem
\bibitem[Choi \latin{et~al.}(2020)Choi, Holehouse, and Pappu]{choi_physical_2020}
Choi,~J.-M.; Holehouse,~A.~S.; Pappu,~R.~V. Physical {Principles} {Underlying} the {Complex} {Biology} of {Intracellular} {Phase} {Transitions}. \emph{Annu. Rev. Biophys.} \textbf{2020}, \emph{49}, 107--133\relax
\mciteBstWouldAddEndPuncttrue
\mciteSetBstMidEndSepPunct{\mcitedefaultmidpunct}
{\mcitedefaultendpunct}{\mcitedefaultseppunct}\relax
\EndOfBibitem
\bibitem[Dietz and Rief(2006)Dietz, and Rief]{dietz_protein_2006}
Dietz,~H.; Rief,~M. Protein structure by mechanical triangulation. \emph{Proc. Natl. Acad. Sci. U. S. A.} \textbf{2006}, \emph{103}, 1244--1247\relax
\mciteBstWouldAddEndPuncttrue
\mciteSetBstMidEndSepPunct{\mcitedefaultmidpunct}
{\mcitedefaultendpunct}{\mcitedefaultseppunct}\relax
\EndOfBibitem
\bibitem[Oesterhelt \latin{et~al.}(2000)Oesterhelt, Oesterhelt, Pfeiffer, Engel, Gaub, and Müller]{oesterhelt_unfolding_2000}
Oesterhelt,~F.; Oesterhelt,~D.; Pfeiffer,~M.; Engel,~A.; Gaub,~H.~E.; Müller,~D.~J. Unfolding {Pathways} of {Individual} {Bacteriorhodopsins}. \emph{Science} \textbf{2000}, \emph{288}, 143--146\relax
\mciteBstWouldAddEndPuncttrue
\mciteSetBstMidEndSepPunct{\mcitedefaultmidpunct}
{\mcitedefaultendpunct}{\mcitedefaultseppunct}\relax
\EndOfBibitem
\bibitem[Lide(1991)]{lide_crc_1991}
Lide,~D.~R. \emph{{CRC} {Handbook} of {Chemistry} and {Physics}}, 72nd ed.; CRC Press: Boca Raton, FL, 1991\relax
\mciteBstWouldAddEndPuncttrue
\mciteSetBstMidEndSepPunct{\mcitedefaultmidpunct}
{\mcitedefaultendpunct}{\mcitedefaultseppunct}\relax
\EndOfBibitem
\bibitem[Fredrickson(2006)]{fredrickson_equilibrium_2006}
Fredrickson,~G.~H. \emph{The equilibrium theory of inhomogeneous polymers}; International series of monographs on physics 134; Clarendon Press; Oxford University Press: Oxford : New York, 2006\relax
\mciteBstWouldAddEndPuncttrue
\mciteSetBstMidEndSepPunct{\mcitedefaultmidpunct}
{\mcitedefaultendpunct}{\mcitedefaultseppunct}\relax
\EndOfBibitem
\bibitem[Duan \latin{et~al.}(2020)Duan, Li, and Wang]{duan_conformation_2020}
Duan,~C.; Li,~W.; Wang,~R. Conformation of a single polyelectrolyte in poor solvents. \emph{J. Chem. Phys.} \textbf{2020}, \emph{153}, 064901\relax
\mciteBstWouldAddEndPuncttrue
\mciteSetBstMidEndSepPunct{\mcitedefaultmidpunct}
{\mcitedefaultendpunct}{\mcitedefaultseppunct}\relax
\EndOfBibitem
\bibitem[Duan and Wang(2022)Duan, and Wang]{duan_association_2022}
Duan,~C.; Wang,~R. Association of two polyelectrolytes in salt solutions. \emph{Soft Matter} \textbf{2022}, \emph{18}, 6934--6941\relax
\mciteBstWouldAddEndPuncttrue
\mciteSetBstMidEndSepPunct{\mcitedefaultmidpunct}
{\mcitedefaultendpunct}{\mcitedefaultseppunct}\relax
\EndOfBibitem
\bibitem[Duan \latin{et~al.}(2022)Duan, Li, and Wang]{duan_stable_2022}
Duan,~C.; Li,~W.; Wang,~R. Stable {Vesicles} {Formed} by a {Single} {Polyelectrolyte} in {Salt} {Solutions}. \emph{Macromolecules} \textbf{2022}, \emph{55}, 906--913\relax
\mciteBstWouldAddEndPuncttrue
\mciteSetBstMidEndSepPunct{\mcitedefaultmidpunct}
{\mcitedefaultendpunct}{\mcitedefaultseppunct}\relax
\EndOfBibitem
\bibitem[Duan and Wang(2023)Duan, and Wang]{duan_electrostatics-induced_2023}
Duan,~C.; Wang,~R. Electrostatics-{Induced} {Nucleated} {Conformational} {Transition} of {Protein} {Aggregation}. \emph{Phys. Rev. Lett.} \textbf{2023}, \emph{130}, 158401\relax
\mciteBstWouldAddEndPuncttrue
\mciteSetBstMidEndSepPunct{\mcitedefaultmidpunct}
{\mcitedefaultendpunct}{\mcitedefaultseppunct}\relax
\EndOfBibitem
\bibitem[Israelachvili(2011)]{israelachvili_intermolecular_2011}
Israelachvili,~J.~N. \emph{Intermolecular and {Surface} {Forces}}, 3rd ed.; Elsevier Academic Press: San Diego, CA, 2011\relax
\mciteBstWouldAddEndPuncttrue
\mciteSetBstMidEndSepPunct{\mcitedefaultmidpunct}
{\mcitedefaultendpunct}{\mcitedefaultseppunct}\relax
\EndOfBibitem
\bibitem[Wang and Wang(2014)Wang, and Wang]{wang_theory_2014}
Wang,~R.; Wang,~Z.-G. Theory of {Polymer} {Chains} in {Poor} {Solvent}: {Single}-{Chain} {Structure}, {Solution} {Thermodynamics}, and {Theta} {Point}. \emph{Macromolecules} \textbf{2014}, \emph{47}, 4094--4102\relax
\mciteBstWouldAddEndPuncttrue
\mciteSetBstMidEndSepPunct{\mcitedefaultmidpunct}
{\mcitedefaultendpunct}{\mcitedefaultseppunct}\relax
\EndOfBibitem
\bibitem[Chantawansri \latin{et~al.}(2011)Chantawansri, Hur, García-Cervera, Ceniceros, and Fredrickson]{chantawansri_spectral_2011}
Chantawansri,~T.~L.; Hur,~S.-M.; García-Cervera,~C.~J.; Ceniceros,~H.~D.; Fredrickson,~G.~H. Spectral collocation methods for polymer brushes. \emph{J. Chem. Phys.} \textbf{2011}, \emph{134}, 244905\relax
\mciteBstWouldAddEndPuncttrue
\mciteSetBstMidEndSepPunct{\mcitedefaultmidpunct}
{\mcitedefaultendpunct}{\mcitedefaultseppunct}\relax
\EndOfBibitem
\bibitem[Hoffman and Frankel(2018)Hoffman, and Frankel]{hoffman_numerical_2018}
Hoffman,~J.~D.; Frankel,~S. \emph{Numerical {Methods} for {Engineers} and {Scientists}}; CRC Press, 2018\relax
\mciteBstWouldAddEndPuncttrue
\mciteSetBstMidEndSepPunct{\mcitedefaultmidpunct}
{\mcitedefaultendpunct}{\mcitedefaultseppunct}\relax
\EndOfBibitem
\bibitem[Klushin \latin{et~al.}(2001)Klushin, Birshtein, and Amoskov]{klushin_microphase_2001}
Klushin,~L.~I.; Birshtein,~T.~M.; Amoskov,~V.~M. Microphase {Coexistence} in {Brushes}. \emph{Macromolecules} \textbf{2001}, \emph{34}, 9156--9167\relax
\mciteBstWouldAddEndPuncttrue
\mciteSetBstMidEndSepPunct{\mcitedefaultmidpunct}
{\mcitedefaultendpunct}{\mcitedefaultseppunct}\relax
\EndOfBibitem
\bibitem[Ballauff and Borisov(2016)Ballauff, and Borisov]{ballauff_phase_2016}
Ballauff,~M.; Borisov,~O. Phase transitions in {Homopolymers}. \emph{Polymer} \textbf{2016}, 402--408\relax
\mciteBstWouldAddEndPuncttrue
\mciteSetBstMidEndSepPunct{\mcitedefaultmidpunct}
{\mcitedefaultendpunct}{\mcitedefaultseppunct}\relax
\EndOfBibitem
\bibitem[Borisov \latin{et~al.}(1994)Borisov, Zhulina, and Birshtein]{borisov_diagram_1994}
Borisov,~O.~V.; Zhulina,~E.~B.; Birshtein,~T.~M. Diagram of the {States} of a {Grafted} {Polyelectrolyte} {Layer}. \emph{Macromolecules} \textbf{1994}, \emph{27}, 4795--4803\relax
\mciteBstWouldAddEndPuncttrue
\mciteSetBstMidEndSepPunct{\mcitedefaultmidpunct}
{\mcitedefaultendpunct}{\mcitedefaultseppunct}\relax
\EndOfBibitem
\bibitem[Skoda(2019)]{skoda_recent_2019}
Skoda,~M. W.~A. Recent developments in the application of {X}-ray and neutron reflectivity to soft-matter systems. \emph{Curr. Opin. Colloid Interface Sci.} \textbf{2019}, \emph{42}, 41--54\relax
\mciteBstWouldAddEndPuncttrue
\mciteSetBstMidEndSepPunct{\mcitedefaultmidpunct}
{\mcitedefaultendpunct}{\mcitedefaultseppunct}\relax
\EndOfBibitem
\bibitem[Xu \latin{et~al.}(2013)Xu, Penfold, Thomas, Petkov, Tucker, and Webster]{xu_formation_2013}
Xu,~H.; Penfold,~J.; Thomas,~R.~K.; Petkov,~J.~T.; Tucker,~I.; Webster,~J. P.~R. The {Formation} of {Surface} {Multilayers} at the {Air}–{Water} {Interface} from {Sodium} {Polyethylene} {Glycol} {Monoalkyl} {Ether} {Sulfate}/{AlCl3} {Solutions}: {The} {Role} of the {Size} of the {Polyethylene} {Oxide} {Group}. \emph{Langmuir} \textbf{2013}, \emph{29}, 11656--11666\relax
\mciteBstWouldAddEndPuncttrue
\mciteSetBstMidEndSepPunct{\mcitedefaultmidpunct}
{\mcitedefaultendpunct}{\mcitedefaultseppunct}\relax
\EndOfBibitem
\bibitem[Russell(1990)]{russell_x-ray_1990}
Russell,~T.~P. X-ray and neutron reflectivity for the investigation of polymers. \emph{Mater. Sci. Rep.} \textbf{1990}, \emph{5}, 171--271\relax
\mciteBstWouldAddEndPuncttrue
\mciteSetBstMidEndSepPunct{\mcitedefaultmidpunct}
{\mcitedefaultendpunct}{\mcitedefaultseppunct}\relax
\EndOfBibitem
\bibitem[Braslau \latin{et~al.}(1985)Braslau, Deutsch, Pershan, Weiss, Als-Nielsen, and Bohr]{braslau_surface_1985}
Braslau,~A.; Deutsch,~M.; Pershan,~P.~S.; Weiss,~A.~H.; Als-Nielsen,~J.; Bohr,~J. Surface {Roughness} of {Water} {Measured} by {X}-{Ray} {Reflectivity}. \emph{Phys. Rev. Lett.} \textbf{1985}, \emph{54}, 4\relax
\mciteBstWouldAddEndPuncttrue
\mciteSetBstMidEndSepPunct{\mcitedefaultmidpunct}
{\mcitedefaultendpunct}{\mcitedefaultseppunct}\relax
\EndOfBibitem
\bibitem[Reinhardt \latin{et~al.}(2013)Reinhardt, Dzubiella, Trapp, Gutfreund, Kreuzer, Gröschel, Müller, Ballauff, and Steitz]{reinhardt_fine-tuning_2013}
Reinhardt,~M.; Dzubiella,~J.; Trapp,~M.; Gutfreund,~P.; Kreuzer,~M.; Gröschel,~A.~H.; Müller,~A. H.~E.; Ballauff,~M.; Steitz,~R. Fine-{Tuning} the {Structure} of {Stimuli}-{Responsive} {Polymer} {Films} by {Hydrostatic} {Pressure} and {Temperature}. \emph{Macromolecules} \textbf{2013}, \emph{46}, 6541--6547\relax
\mciteBstWouldAddEndPuncttrue
\mciteSetBstMidEndSepPunct{\mcitedefaultmidpunct}
{\mcitedefaultendpunct}{\mcitedefaultseppunct}\relax
\EndOfBibitem
\bibitem[Laloyaux \latin{et~al.}(2010)Laloyaux, Mathy, Nysten, and Jonas]{laloyaux_surface_2010}
Laloyaux,~X.; Mathy,~B.; Nysten,~B.; Jonas,~A.~M. Surface and {Bulk} {Collapse} {Transitions} of {Thermoresponsive} {Polymer} {Brushes}. \emph{Langmuir} \textbf{2010}, \emph{26}, 838--847\relax
\mciteBstWouldAddEndPuncttrue
\mciteSetBstMidEndSepPunct{\mcitedefaultmidpunct}
{\mcitedefaultendpunct}{\mcitedefaultseppunct}\relax
\EndOfBibitem
\bibitem[Yim \latin{et~al.}(2005)Yim, Kent, Satija, Mendez, Balamurugan, Balamurugan, and Lopez]{yim_evidence_2005}
Yim,~H.; Kent,~M.~S.; Satija,~S.; Mendez,~S.; Balamurugan,~S.~S.; Balamurugan,~S.; Lopez,~G.~P. Evidence for vertical phase separation in densely grafted, high-molecular-weight poly(N-isopropylacrylamide) brushes in water. \emph{Phys. Rev. E} \textbf{2005}, \emph{72}, 051801\relax
\mciteBstWouldAddEndPuncttrue
\mciteSetBstMidEndSepPunct{\mcitedefaultmidpunct}
{\mcitedefaultendpunct}{\mcitedefaultseppunct}\relax
\EndOfBibitem
\bibitem[Balastre \latin{et~al.}(2002)Balastre, Li, Schorr, Yang, Mays, and Tirrell]{balastre_study_2002}
Balastre,~M.; Li,~F.; Schorr,~P.; Yang,~J.; Mays,~J.~W.; Tirrell,~M.~V. A {Study} of {Polyelectrolyte} {Brushes} {Formed} from {Adsorption} of {Amphiphilic} {Diblock} {Copolymers} {Using} the {Surface} {Forces} {Apparatus}. \emph{Macromolecules} \textbf{2002}, \emph{35}, 9480--9486\relax
\mciteBstWouldAddEndPuncttrue
\mciteSetBstMidEndSepPunct{\mcitedefaultmidpunct}
{\mcitedefaultendpunct}{\mcitedefaultseppunct}\relax
\EndOfBibitem
\bibitem[Drobek \latin{et~al.}(2005)Drobek, Spencer, and Heuberger]{drobek_compressing_2005}
Drobek,~T.; Spencer,~N.~D.; Heuberger,~M. Compressing {PEG} {Brushes}. \emph{Macromolecules} \textbf{2005}, \emph{38}, 5254--5259\relax
\mciteBstWouldAddEndPuncttrue
\mciteSetBstMidEndSepPunct{\mcitedefaultmidpunct}
{\mcitedefaultendpunct}{\mcitedefaultseppunct}\relax
\EndOfBibitem
\bibitem[Balzer and Wang(2023)Balzer, and Wang]{balzer_electroresponse_2023}
Balzer,~C.; Wang,~Z.-G. Electroresponse of weak polyelectrolyte brushes. \emph{Eur. Phys. J. E: Soft Matter Biol. Phys.} \textbf{2023}, \emph{46}, 82\relax
\mciteBstWouldAddEndPuncttrue
\mciteSetBstMidEndSepPunct{\mcitedefaultmidpunct}
{\mcitedefaultendpunct}{\mcitedefaultseppunct}\relax
\EndOfBibitem
\end{mcitethebibliography}

\onecolumn
\subsection*{\Large{Supplemental Information}}
\bigskip
\renewcommand{\theequation}{S\arabic{equation}}
\setcounter{equation}{0}
\renewcommand{\thefigure}{S\arabic{figure}}
\setcounter{figure}{0}
\renewcommand{\thetable}{S\arabic{table}}
\setcounter{table}{0}

\subsection{I. NFH Sequence and Model Parameters for Amino Acids}
In this section, the amino acid sequence of NFH and the amino acid parameters used in this work are provided.

The NFH sequence was obtained from the authors of Srinivasan et al. \cite{srinivasan_stimuli-sensitive_2014}:
\seqsplit{MGCWYMSEFTSMSTHIKVKSEEKIKVVEKSEKETVIVEEQTEEIQVTEEVTEEEDKEAQGEEEEEAEEGGEEAATTSPPA
EEAASPEKETKSPVKEEAKSPAEAKSPAEAKSPAEAKSPAEVKSPAVAKSPAEVKSPAEVKSPAEAKSPAEAKSPAEVKSPATVKSPG
EAKSPAEAKSPAEVKSPVEAKSPAEAKSPASVKSPGEAKSPAEAKSPAEVKSPATVKSPVEAKSPAEVKSPVTVKSPAEAKSPVEVKSP
ASVKSPSEAKSPAGAKSPAEAKSPVVAKSPAEAKSPAEAKPPAEAKSPAEAKSPAEAKSPAEAKSPAEAKSPVEVKSPEKAKSPVKEG
AKSLAEAKSPEKAKSPVKEEIKPPAEVKSPEKAKSPMKEEAKSPEKAKTLDVKSPEAKTPAKEEAKRPADIRSPEQVKSPAKEEAKSP
EKEETRTEKVAPKKEEVKSPVEEVKAKEPPKKVEEEKTPATPKTEVKESKKDEAPKEAQKPKAEEKEPLTEKPKDSPGEAK
KEEAKEKKAAAPEEETPAKLGVKEEAKPKEKAEDAKAKEPSKPSEKEKPKKEEVPAAPEKKDTKEEKTTESKKPEEKPKME
AKAKEEDKGLPQEPSKPKTEKAEKSSSTDQKDSQPSEKAPEDKLLEHHHHHH}\hfill\par
The interaction parameter $\chi_i$ between block $i$ and solvent are calculated by the average of the interaction parameters of its constituent amino acids. The interaction parameters between amino acids and solvent are based on groups and values previously used in the literature \cite{zhulina_effect_2007, lee_effects_2013}:

\indent 
\{G, P, C, M, A, L, V, I\} = 4.0

\indent
\{Y, Q, N, H, F, W, S, T\} = 0.6

\indent
\{E, D, K, R\} = 0.0 

\bigskip
The charge on each amino acid at the bulk pH is calculated using the Henderson--Hasselbalch equation and the dissociation constants for acidic and basic residues \cite{lide_crc_1991}:
\break

\noindent
$\mathrm{pK_a}$:

\indent
D: 3.65, E: 4.25

\noindent
$\mathrm{pK_b}$:

\indent
H: 6.00, K: 10.53, R: 12.48, C: 8.18, Y: 10.07

\clearpage
\subsection{II. Derivation of Self-consistent Field Theory for Protein Brushes}
In this section, a detailed derivation of the key equations in the SCFT for the multiblock macromolecular model used for protein brushes is provided. In Subsection II.1, the derivation for proteins with quenched charge is provided, which is directly calculated from bulk pH. In Subsection II.2, an annealed model is adopted for the protein charge, which explicitly considers the local proton concentration.

\subsubsection{II.1. Quenched Charge Model for Proteins}
The particle-based partition function and Hamiltonian are given in Eqs.~1 and 2 of the main text. The transform from particle-based to field-based representation is conducted through the identity transform: 

\begin{equation}\label{Seqn:fourier}
1 = \int \mathcal{D}\phi_k\ \prod_{\br} \delta \big[\phi_k(\br) - \hat{\phi}_k(\br)\big] = \int \mathcal{D}\phi_k\mathcal{D}w_k \exp \bigg\{i \int \dr w_k(\br)\big[\phi_k(\br)-\hat{\phi}_k(\br)\big]\bigg\}\quad k=(i, s)\\
\end{equation}
These transforms introduce the Fourier conjugate fields $w_i$ and $w_s$ for each protein block $i$ and solvent, respectively. 
The Fourier representation of the incompressibility condition is
\begin{equation}\label{Seqn:eta}
1 = \int \mathcal{D}\eta\ \prod_{\br} \delta \bigg[1 - \sum_{i=1}^{\xi}\hat{\phi}_i(\br) - \hat{\phi}_s(\br)\bigg] = \int \mathcal{D}\eta \exp \bigg\{i \int \dr \eta(\br)\bigg[1-\sum_{i=1}^{\xi}\hat{\phi}_i(\br)-\hat{\phi}_s(\br)\bigg]\bigg\}
\end{equation}
where $\eta$ is the conjugate field. Substituting Eqs.~\ref{Seqn:fourier} and \ref{Seqn:eta} into the partition function yields:
\begin{equation}\begin{split} \label{Seqn:Xi}
\Xi = &\int \prod_{i=1}^{\xi} \mathcal{D}\phi_i\mathcal{D}w_i \mathcal{D}\phi_s \mathcal{D}w_s\mathcal{D}\eta\ \frac{1}{n_p!\nu^{N n_p}} \prod_{\gamma} \sum_{n_\gamma=0}^{\infty} \frac{e^{\mu_\gamma n_\gamma}}{n_\gamma!\nu_\gamma^{n_\gamma}} \\
&\int \prod_{j=1}^{n_\gamma}\dr[\gamma, j] \int \prod_{k=1}^{n_p}\mathcal{D}\{\bR_k\}\ \exp \bigg\{ - \frac{3}{2b^2}\int_{0}^{N} \dvar{s}\bigg(\frac{\partial \bR_k (\textrm{s})}{\partial \textrm{s}}\bigg)^2\bigg\}\\
&\exp\bigg\{ - \frac{1}{\nu}\int\dr \sum_{i=1}^\xi \bigg(\chi_i \phi_i\phi_s + \sum_{j\ne i}^\xi \frac{\chi_{ij}}{2} \phi_i\phi_j\bigg) -\frac{\beta e^2}{2} \int\dr\drsup[\prime] \hat{\rho}_c(\br)\ C(\br, \brsup[\prime])\ \hat{\rho}_c(\brsup[\prime])\bigg\}\\
&\exp\bigg\{\frac{1}{\nu}\int \dr \bigg[ i\eta \big(\sum_{i=1}^{\xi}\phi_i + \phi_s - 1\big) + i \sum_{i=1}^{\xi} \big(w_i \big(\phi_i - \hat{\phi}_i\big)\big) + iw_s \big(\phi_s - \hat{\phi}_s \big)\bigg]\bigg\}
~~ (\gamma = s, \pm)
\end{split}\end{equation}
Note that we assume $\nu_p=\nu_s=\nu$.

By grouping polymer-specific and solvent-specific terms, we arrive at the single-particle solvent partition function $Q_s$ and single-chain polymer partition function $Q_p$:
\begin{equation}\begin{split}
Q_s &= \frac{1}{\nu} \int \mathrm{d}\mathbf{r}_\beta\ \exp \big(-iw_s(\mathbf{r}_\beta)\big)\\
Q_p &= \frac{1}{\nu^N}\int\mathcal{D}\bR\ \exp \left\{-\int_{0}^{N} \dvar{s} \bigg[\frac{3}{2b^2}\bigg(\frac{\partial \bR}{\partial \mathrm{s}}\bigg)^2-iw_i(\bR)\bigg]\right\}\\
&= \frac{1}{\nu} \int \dr q(\br; N)\ ,
\end{split}\end{equation}
where the propagator $q(\br; \mathrm{s})$ originating from the grafted end in the final line satisfies the modified diffusion equation (Eq. 4 in the main text)\cite{fredrickson_equilibrium_2006} : 

\begin{align}\label{Seqn:MDE}
\frac{\partial}{\partial s} q(\br;\mathrm{s}) = \frac{b^2}{6}&\nabla^2 q(\br;s)- w_i(\br)q(\br;s);,\\
\text{where}\; w_i(\br) &= \nonumber
\begin{cases}
w_{1}(\br) & \text{for } s = [0, N_1]\\
&\vdots \\
w_{\xi}(\br) & \text{for } s = [\sum_{j=0}^{\xi-1}N_j, \sum_{j=0}^{\xi} N_j]
\end{cases}
\end{align}
The counter propagator $q_c(\br; \mathrm{s})$ originating from the free end also satisfies the modified diffusion equation.
Finally, we perform the Hubbard--Stratonovich transformation to decouple the interactions between charged particles: 

\begin{equation}\begin{split}
\exp \bigg\{ -\frac{\beta}{2} &\int \mathrm{d}\mathbf{r}\mathrm{d}\mathbf{r}^\prime\ \hat{\rho}_c(\mathbf{r})C(\mathbf{r}, \mathbf{r}^\prime)\hat\rho_c(\mathbf{r}^\prime) \bigg\}\\ 
&= C_\psi\int\mathcal{D}\psi\ \exp \bigg\{ -\frac{\beta}{2}\int\mathrm{d}\mathbf{r}\mathrm{d}\mathbf{r}^\prime\ \psi(\mathbf{r})C^{-1}(\mathbf{r}, \mathbf{r}^\prime)\psi(\mathbf{r}^\prime) -i\beta\int \mathrm{d}\mathbf{r}\ \hat{\rho}_c(\mathbf{r})\psi(\mathbf{r})\bigg\}\\
&= C_\psi \int\mathcal{D}\psi\ \exp \bigg\{\int \mathrm{d}\mathbf{r}\ \bigg[\frac{\epsilon(\br)}{2}\lvert i\psi(\br)\rvert^2- \hat{\rho}_c i \psi(\mathbf{r}) \bigg]\bigg\}\ ,
\end{split}\end{equation}
where in the last step, the dielectric constant and electrostatic potential were made dimensionless: $\epsilon = \epsilon/(\beta e^2)$, $\psi=\beta e\psi$. 
$C^{-1}(\mathbf{r}, \mathbf{r}^\prime) = -\nabla\cdot\big[\epsilon(\mathbf{r})\nabla\big]\delta(\mathbf{r}-\mathbf{r}^\prime)$ is the functional inverse of the Coulomb operator, $C(\mathbf{r},\mathbf{r}^\prime)$. The normalization constant, $C_{\psi}^{-1} = \int\mathcal{D}\psi\ \exp \big\{ -(\beta/2)\int\mathrm{d}\mathbf{r}\mathrm{d}\mathbf{r}^\prime\ \psi(\mathbf{r})C^{-1}(\mathbf{r}, \mathbf{r}^\prime)\psi(\mathbf{r}^\prime)\big\}$, is discarded as it will only shift the partition function by a constant. 

The final partition function is thus:

\begin{equation}\begin{split}
\Xi = &\int \prod_{i=1}^{\xi} \mathcal{D}\phi_i\mathcal{D}w_i \mathcal{D}\phi_s \mathcal{D}w_s\mathcal{D}\eta\mathcal{D}\psi\ \exp(e^{\beta\mu_s} Q_s^{n_s})\  \frac{Q_p^{n_p}}{n_p!}\\
&\exp\bigg\{ \frac{1}{\nu} \int \dr \bigg[ \sum_{i=1}^\xi \big(\chi_i \phi_i\phi_s + \sum_{j\ne i}^\xi \frac{\chi_{ij}}{2} \phi_i\phi_j\big) + \sum_{i=1}^{\xi}i w_i\phi_i + iw_s\phi_s+i\eta\big(\phi_p+\phi_s-1\big)\bigg]\bigg \}\\
&\exp\bigg\{\frac{1}{\nu}\int\dr \bigg[  \frac{\epsilon}{2}\lvert i \psi \rvert^2 + \lambda_\pm \exp\big(\mp z_\pm i \psi\big)+i\psi\sum_{i=1}^{\xi} \frac{\alpha_i}{\nu} \phi_i \bigg]\bigg\}\ ,
\end{split}\end{equation}
where $\phi_p(\br) = \sum_{i=1}^{\xi} \phi_i(\br)$ is the total protein density and $\lambda_\pm = e^{\mu_\pm}/\nu_\pm$ is the fugacity of the ions.
Here, we consider a one-dimensional planar system, where the protein densities vary only in the $z$-direction but remain homogeneous in the $xy$-plane.
After applying the saddle-point approximation and replacing $i\Psi \to \Psi$ for the purely imaginary saddle-point values in the fields $\Psi = \phi_i, w_i, \phi_s, w_s, \eta, \psi$, we arrive at the equilibrium free energy per unit area (Eq. 5 in the main text):

\begin{align}
F =&-\sigma \ln Q_p - e^{\mu_s}Q_s\nonumber\\ &+\frac{1}{\nu}\int \dvar{z} \bigg[ \sum_{i=1}^{\xi} \big(\chi_i\phi_i\phi_s - w_i\phi_i\big) - w_s\phi_s - \eta\big(\phi_p+\phi_s-1\big) \bigg]\nonumber\\ &+ \int \dvar{z} \bigg[-\frac{\epsilon}{2}\lvert \nabla\psi \rvert^2 +\frac{\psi}{\nu}\sum_{i=1}^{\xi}\alpha_i\phi_i - c_+-c_-\bigg]\ , \label{Seqn:F}
\end{align}
where $c_\pm = \lambda_\pm\exp(\mp z_\pm\psi)$ is the ion concentration and $\chi_{ij}=0$ is assumed. Thus, $\sigma = n_p / A$ is the grafting density of the protein brush, where $A$ is the area of the plate. 
After functional minimization of $F$ with respect to all the fields (i.e., $\phi_i, w_i, w_s, \eta, \psi$) in addition to assuming that $\epsilon(\mathrm{z})$ can be expressed as a function of $\phi_i(\mathrm{z})$, we obtain the following coupled, self-consistent field equations:

\begin{subequations} \label{SSCF:main}
\begin{align}
w_i(\mathrm{z}) &= \chi_i\phi_s(\mathrm{z}) - \alpha_i\psi(\mathrm{z}) -\frac{\nu}{2}\frac{\partial \epsilon(\mathrm{z})}{\partial \phi_i} \bigg\lvert \frac{d\psi(\mathrm{z})}{d\mathrm{z}}\bigg\rvert^2 - \eta(\mathrm{z}) \label{SSCF:a} \\ 
w_s(\mathrm{z}) &= \sum_{i=1}^{\xi}\chi_i\phi_i(\mathrm{z}) - \eta(\mathrm{z}) \label{SSCF:b} \\
\phi_{i}(\mathrm{z}) &= \frac{\sigma}{Q_p} \int_{\sum_{j=0}^{i-1}N_{j}}^{\sum_{j=0}^{i}N_j} \mathrm{ds}\ q(\mathrm{z};\mathrm{s}) q_c(\mathrm{z};\mathrm{s})\label{SSCF:c}\\
\phi_s(\mathrm{z}) &= e^{\mu_s}\exp\big(-w_s(\mathrm{z})\big)\label{SSCF:d}\\
-\frac{d}{d\mathrm{z}} \bigg(\epsilon(\mathrm{z})\frac{d}{d\mathrm{z}}\psi(\mathrm{z}) \bigg) &= z_+c_+(\mathrm{z})-z_-c_-(\mathrm{z}) +\sum_{i=1}^{\xi}\frac{\alpha_i}{\nu} \phi_i(\mathrm{z})\label{SSCF:e}
\end{align}
\end{subequations}

For the Poisson-Boltzmann Equation (Eq.~\ref{SSCF:e}), $d\psi/d\mathrm{z} = 0$ and $\psi(\mathrm{z}) = 0$ are used for the boundary conditions at $\mathrm{z}=0$ and $\mathrm{z}=\infty$, respectively. For the modified diffusion equation (Eq.~\ref{Seqn:MDE}), $q(\mathrm{z}; s) = 0$ is used at both boundaries. The corresponding boundary conditions for $q_c(\mathrm{z}; s)$ are identical to $q(\mathrm{z}; s)$. However, the initial conditions are not: $q(\mathrm{z}; 0) = \delta(\mathrm{z}-\mathrm{z}^*)$ with $z^* \to 0_+$ and $q_c(\mathrm{z}; N) = 1$.

The charge densities of amino acids at bulk pH, $\alpha_b^*$, are related to the dissociation constants $\mathrm{pK_a}$ and $\mathrm{pK_b}$ for the acidic and basic residues, respectively. From the Henderson-Hasselbalch equation, $\log_{10}r_+^b = -(\mathrm{pH-pK_b})$, where $r_+^b$ is the ratio of the number of positively charged residues to the number of uncharged residues at bulk conditions. Similarly, $\log_{10}r_-^b = \mathrm{pH-pK_a}$ for the ratio of negatively charged to uncharged residues. Thus, for an arbitrary amino acid, the bulk charge density can be calculated using the following equation:
\begin{equation}
    \alpha_b^* = \frac{r_+^b}{1+r_+^b} - \frac{r_-^b}{1+r_-^b} \label{Seqn:alphaHH}
\end{equation}
The charge density $\alpha_i$ of block $i$ can then be calculated by adding the charge densities of its constituent amino acids. 

\subsubsection{II.2. Annealed Charge Model for Proteins}
Our theory can also incorporate an explicit treatment of the local proton concentration for a more rigorous treatment of the residue dissociation reaction. This is especially important at high pH, at which the hydrogen concentration in the brush may vary significantly from the bulk value. To illustrate the methodology, we summarize here the derivation for a simplified case in which all the residues are basic with the same value of $\mathrm{pK_b}$. It can be easily generalized to model real proteins where the dissociation constant of each residue is different. A general idea is to modify the previous quenched model by the following two aspects: (1) the chemical distinguishability of \ch{H+} and \ch{OH-} from salt ions ($\pm$) and the new corresponding chemical potentials $\mu_{\ch{H+}}$ and $\mu_{\ch{OH-}}$; (2) the introduction of the microscopic protein charge density operator. Taking a polybase with the same $\mathrm{pK_b}$ for all the residues as an example, the charge density operator $\hat{\alpha}(\br)$ can be written as
\begin{equation}
    \hat{\alpha}(\br) = \frac{\sum_{C=1}^{n_C} \delta(\br-\br[C])}{\sum_{k=1}^{n_p} \int_0^N \delta(\br-\bR_k(s))}~,
\end{equation}
where $\br[C]$ are the locations of each of the $n_C$ charged segments. A similar operator for the microscopic protein uncharged fraction operator $\hat{\alpha}_0$ is also introduced. To specify $\hat{\alpha}$ as the amount of charge from the dissociated residues, the delta function $\prod_{\br}\delta[ \hat{\alpha}(\br)\hat{\phi}_p(\br) + \hat{\alpha}_0\hat{\phi}_p(\br) - \hat{\phi}_p(\br)]$ is incorporated into the partition function.

After following the standard self-consistent field procedure as described at the beginning of this section, we arrive at a set of self-consistent field equations similar to Eqs.~\ref{SSCF:main}:
\begin{align}
    w_p(\mathrm{z}) &= \chi\phi_s(\mathrm{z}) + \eta(\mathrm{z})\\ 
    w_s(\mathrm{z}) &= \chi\phi_p(\mathrm{z}) + \eta(\mathrm{z})\\
    \phi_p(\mathrm{z}) &= \frac{\sigma}{Q_p} \int_0^N \dvar{s} q(\mathrm{z}; \mathrm{s}) q_c(\mathrm{z}; \mathrm{s})\\
    \phi_s(\mathrm{z}) &= e^\mu_s \exp\big(-w_s(\mathrm{z})\big)\\
    -\frac{d}{dz}\bigg(\epsilon(\mathrm{z})\frac{d}{dz}\psi(\mathrm{z})\bigg) &= \sum_{\gamma} Z_\gamma c_\gamma(\mathrm{z}) + \frac{\alpha(\mathrm{z})}{\nu} \phi_p(\mathrm{z})~,
\end{align}
where $\gamma = \pm, \ch{H+}, \ch{OH-}$. $Z_\gamma$ is the charge of each ion. $c_\gamma(\mathrm{z})= \lambda_\gamma\exp\big(-Z_\gamma\psi(\mathrm{z})\big)$ is the ion concentration, where $\lambda_\gamma = e^{\mu_\gamma}/\nu_\gamma$ is the fugacity of the ions determined by the bulk ion concentration $c_\pm^b$ for salt ions and the bulk pH for \ch{H+} and \ch{OH-}.
The local charge density $\alpha$ of the polybase brush is related to its value in the bulk $\alpha_b = r_+^b / (1 + r_+^b)$ as:
\begin{equation}
    \alpha(\mathrm{z}) = \frac{\alpha_be^{-\psi(\mathrm{z})}}{1 - \alpha_b + \alpha_b e^{-\psi(\mathrm{z})}}
\end{equation}
The modified diffusion equation is then:
\begin{equation}\label{Seqn:MDEzeta}
\frac{\partial}{\partial s} q(\mathrm{z};\mathrm{s}) = \frac{b^2}{6}\frac{d^2}{dz^2} q(\mathrm{z};s)- \big(w_p(\mathrm{z})+\zeta(\mathrm{z}) -1\big)q(\mathrm{z};s)\;,
\end{equation}
where $\zeta(\mathrm{z}) = -\ln \big( 1 - \alpha_b + \alpha_b e^{-\psi(\mathrm{z})}\big)$.

\clearpage
\subsection{III. Sensitivity Analysis of the Coarse-graining Procedure}
In this section, the sensitivity analysis of the coarse-graining approach used in this work is provided. As shown in Fig.~\ref{fig:sens} below, after a threshold degree of coarse-graining (i.e., $N>1$ here), enough chemical information is included and the height response is not significantly affected by the number of blocks used to represent the protein. In this work, we use $N=5$.

\begin{figure}[ht!]
    \centering
    \includegraphics[width=0.50\linewidth]{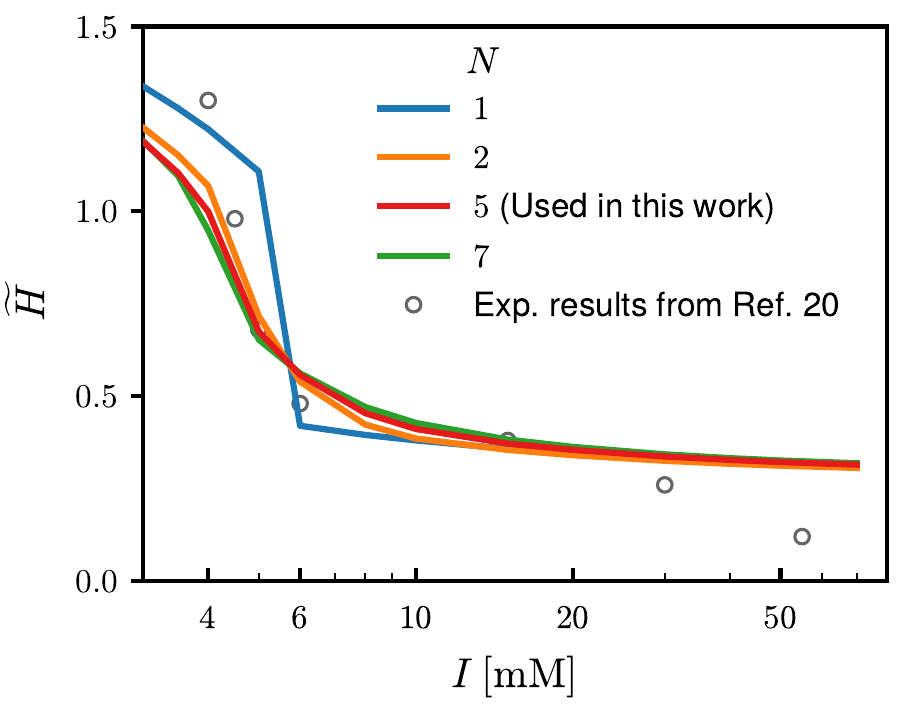} 
    \caption{Height responses to varying ionic strengths at pH 2.4 of brushes comprised of NFH proteins represented by $N = 1,2,5,$ and $7$ number of blocks in the multi-block charged macromolecular model.}
    \label{fig:sens}
\end{figure}

The charge density $\alpha_i$, and hydrophobicity $\chi_i$ of coarse-grained blocks corresponding to $N=1,2,5,$ and $7$ are provided in Table~\ref{tab:NFHtot}.

\begin{table}[t!]
\caption{Partitions of amino acid residues, charge density $\alpha_i$, and hydrophobicity $\chi_i$ of coarse-grained blocks for NFH at pH 2.4 using $N=1, 2, 5, $ and $7$ number of blocks in the multi-block charged macromolecular model.}
\label{tab:NFHtot}
\centering
    \begin{tabular}{c c l c c}
    \hline\hline
    \noalign{\smallskip} 
    $N$         & $i$ & Residues  & $\alpha_i$ & $\chi_i$\\
    \hline \noalign{\smallskip}
    \textbf{1}  & 1 & $[1,~647]$ & 0.209586 & 1.703555\\
    \noalign{\smallskip} \hline \noalign{\smallskip}
    \textbf{2}  & 1 & $[1,~~~~299]$  & 0.147222 & 1.939333\\
                & 2 & $[300,~647]$ & 0.263503 & 1.499712\\
    \noalign{\smallskip} \hline \noalign{\smallskip}
    \textbf{5}  & 1 & $[1,~~~~~28]$ & 0.204967 & 1.586207\\
                & 2 & $[29,~~~~87]$ & 0.027801 & 1.434483\\
                & 3 & $[88,~~319]$ & 0.170493 & 2.113793\\
                & 4 & $[320,~609]$ & 0.261110 & 1.534483\\
                & 5 & $[610,~647]$ & 0.336030 & 0.989474\\
    \noalign{\smallskip} \hline \noalign{\smallskip}
    \textbf{7}  & 1 & $[1,~~~~~31]$ & 0.216566 & 1.456250\\
                & 2 & $[32,~~~~95]$ & 0.056227 & 1.434375\\
                & 3 & $[96,~~319]$& 0.163313 & 2.166071\\
                & 4 & $[320,~511]$& 0.251197 & 1.584375\\
                & 5 & $[512,~543]$& 0.277333 & 1.643750\\
                & 6 & $[544,~607]$& 0.290897 & 1.253125\\
                & 7 & $[608,~647]$& 0.327414 & 1.066667\\
    \noalign{\smallskip}
    \hline\hline
    \end{tabular}
\end{table}

\clearpage
\subsection{IV. Ionic Density Profiles Near the Substrate}
\begin{figure}[ht!]
    \centering
    \includegraphics[width=0.5\linewidth]{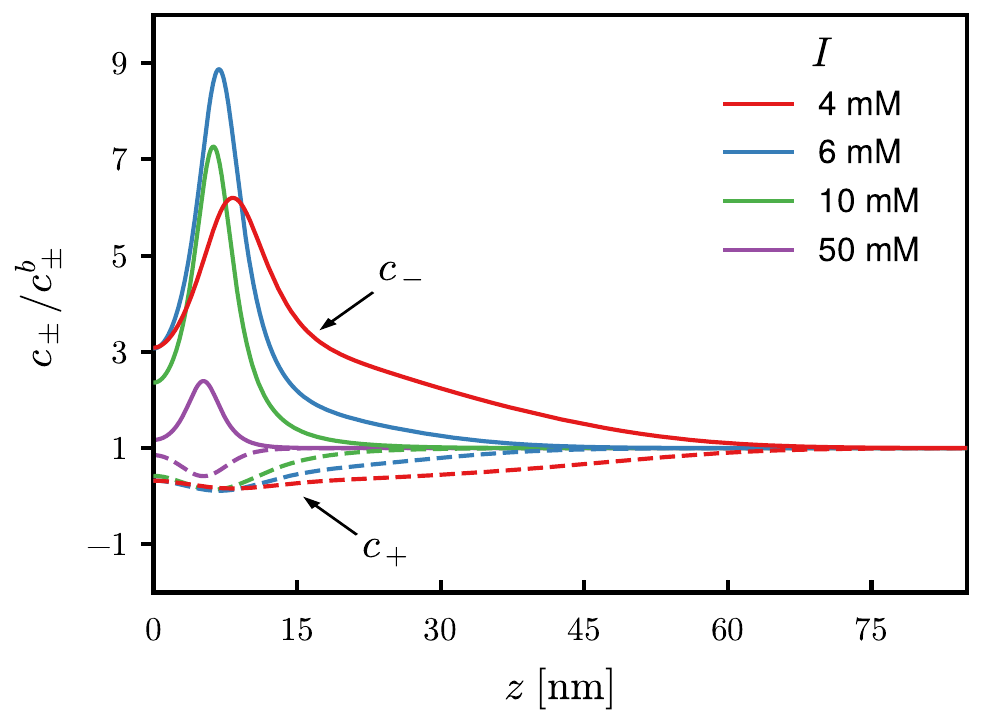} 
    \caption{Representative ionic density profiles $c_\pm/c_\pm^b$ for NFH brushes, where the bulk concentration $c_\pm^b = 4, 6, 10, 50$ mM. $z$ denotes the direction perpendicular to the surface. Anion concentration $c_-$ is shown in solid lines while cation concentration $c_+$ is shown in dashed lines.}
    \label{fig:ions}
\end{figure}

In this section, representative ionic density profiles are provided. Fig.~\ref{fig:ions} shows the ionic density profiles normalized by the bulk ionic salt concentration $c_\pm^b$. Comparing to the representative density profiles in Fig.~3 of the main text, the electric double layer is contained within the brush at high ionic strengths but extends outside of the brush at low ionic strengths. 

\end{document}